\documentclass[dvipsnames,twocolumn]{aastex63}

\usepackage[T1]{fontenc}
\usepackage{nicefrac}
\usepackage{savesym}
\savesymbol{tablenum}
\usepackage{siunitx}
\usepackage{amsmath}
\usepackage{mathtools}
\usepackage{booktabs}
\usepackage{multirow}
\usepackage[normalem]{ulem}
\usepackage{natbib}
\usepackage{dcolumn}
\usepackage{xcolor}
\usepackage{tikz}
\usetikzlibrary{shapes.geometric, arrows}
\usepackage{tablefootnote}

\restoresymbol{SIX}{tablenum}
\DeclareSIUnit\parsec{pc}
\DeclareSIUnit\years{yr}
\DeclareSIUnit\Msol{M_{\odot}}
\DeclareSIUnit\Lsol{L_{\odot}}
\DeclareSIUnit\AU{au}
\DeclareSIUnit\om{\Omega}
\DeclareSIUnit\orb{T_{\mathrm{orb}}}
\DeclareSIUnit\scaleheight{H}
\newcommand{\expnum}[1]{\num[scientific-notation=true, exponent-product=\times, round-mode=places, round-precision=2]{#1}}

\definecolor{dodgerblue}{rgb}{0.11764706, 0.56470588, 1.}
\definecolor{seagreen}{rgb}{0.18039216, 0.54509804, 0.34117647}
\definecolor{maroon}{rgb}{0.50196078, 0., 0.}

\shortauthors{Pfeil et al.}

\newcommand{\rev}[1]{{#1}}
\newcommand{\hubert}[1]{{#1}}
\newcommand{\report}[1]{{{#1}}}
\newcommand{\reporttwo}[1]{{{#1}}}

\newcommand{\rhog}{\rho_{\mathrm{g}}}
\newcommand{\rhod}{\rho_{\mathrm{d}}}
\newcommand{\vg}{\vec{v}_{\mathrm{g}}}

\newcommand{\amax}{a_{\mathrm{max}}}
\newcommand{\amin}{a_{\mathrm{min}}}

\newcommand{\St}{\mathrm{St}}

\newcommand{\der}[2]{\frac{\partial {#1}}{\partial {#2}}}

\newcommand{\planck}{\kappa_{\mathrm{P}}}

\newcommand{\dpy}{{\normalfont\texttt{DustPy}}}
\newcommand{\pluto}{{\normalfont\texttt{PLUTO}}}
\newcommand{\radmc}{{\normalfont\texttt{RADMC-3D}}}

\begin{document}

\title{Dust Coagulation Reconciles Protoplanetary Disk Observations with the Vertical Shear Instability. \\I. Dust Coagulation and the VSI Dead Zone}
\shorttitle{On the Impact of Dust Coagulation on the VSI}

\author[0000-0002-4171-7302]{Thomas Pfeil}
\affiliation{\centering University Observatory, Faculty of Physics, Ludwig-Maximilians-Universität München, Scheinerstr. 1, D-81679 Munich, Germany}
\affiliation{\centering Max-Planck-Institut f\"ur Astronomie, K\"onigstuhl 17, D-69117 Heidelberg, Germany}
\correspondingauthor{Thomas Pfeil}
\email{tpfeil@usm.lmu.de}

\author[0000-0002-1899-8783]{Tilman Birnstiel}
\affiliation{\centering University Observatory, Faculty of Physics, Ludwig-Maximilians-Universität München, Scheinerstr. 1, D-81679 Munich, Germany}
\affiliation{\centering Exzellenzcluster ORIGINS, Boltzmannstr. 2, D-85748 Garching, Germany}

\author[0000-0002-8227-5467]{Hubert Klahr}
\affiliation{\centering Max-Planck-Institut f\"ur Astronomie, K\"onigstuhl 17, D-69117 Heidelberg, Germany}

\begin{abstract}
Protoplanetary disks exhibit a vertical gradient in angular momentum, rendering them susceptible to the Vertical Shear Instability (VSI). The most important condition for the onset of this mechanism is a short timescale of thermal relaxation ($\lesssim 0.1$ orbital timescales). Simulations of fully VSI active disks are characterized by turbulent, vertically extended dust layers. This is in contradiction with recent observations of the outer regions of some protoplanetary disks, which appear highly settled.
In this work, we demonstrate that the process of dust coagulation can diminish the cooling rate of the gas in the outer disk and extinct the VSI activity.
Our findings indicate that the turbulence strength is especially susceptible to variations in the fragmentation velocity of the grains.
A small fragmentation velocity of \report{$\approx$}\SI{100}{\centi \meter \per \second} results in a fully turbulent simulation, whereas a value of \report{$\approx$}\SI{400}{\centi \meter \per \second} results in a laminar outer disk, being consistent with observations.
We show that VSI turbulence remains relatively unaffected by variations in the maximum particle size in the inner disk regions. However, we find that dust coagulation can significantly suppress the occurrence of VSI turbulence at larger distances from the central star.
\end{abstract}


\section{Introduction} \label{sec:intro}
\rev{Around} $\SI{1}{\percent}$ of the mass of protoplanetary disks is \rev{initially} composed of \rev{solids} \citep{Lodders2003, Magg2022}. Despite its small contribution to the overall mass budget, this dust is the building material for planetesimals and planets and an essential observable for infrared and radio observations. It can have a considerable influence on the gas dynamics within the disk via drag forces \citep{Weidenschilling1980, Youdin2005} and is the main source of opacity.
Therefore, cooling and heating are mostly determined by the solids for the bulk of the disk \citep{Semenov2003, Woitke2015, Malygin2017}. 
Many linear instabilities of the gas flow depend on the local rate of thermal relaxation \citep{Klahr2003, Petersen2007a,Petersen2007b,Klahr2014, Lin2015, Marcus2015, Lyra2019} \rev{or the ionization state of the gas \citep{Balbus1991, Blaes1994}}, and are therefore sensitive to the assumed dust size distribution \citep{Barranco2018, Fukuhara2021, Kawasaki2023}. 

In this work, we are specifically interested in the evolution of the vertical shear instability \citep[VSI, ][]{Urpin1998}, which requires a short thermal relaxation time of the gas \citep{Lin2015, Manger2021,Fukuhara2021}. VSI was studied in much detail in isothermal and adiabatic disk models at various rates of $\beta$ cooling \citep[e.g.,][]{Nelson2013} and in models with radiative transfer \citep[e.g.,][]{Stoll2016, Stoll2017, Flores-Rivera2020}. Due to the numerical obstacles of incorporating dust evolution models in hydrodynamic simulations \citep{Drazkowska2014, Gonzalez2017, Drazkowska2019, Lombart2022}, most previous studies consider a static dust population, perfectly coupled to the gas. 
These studies often aim for a detailed analysis of the instability mechanism itself \citep[e.g.,][]{Nelson2013,Manger2021,Svanberg2022}. They showed the VSI's ability to cause large-scale vortex formation \citep{Richard2016, Manger2018, Pfeil2021} and strong corrugations in the dust layer \citep{Stoll2016, Flores-Rivera2020}.  
Simulations assuming perfectly coupled dust or isothermal conditions cannot, however, model the conditions in real protoplanetary disks, for which observations show an evolved dust population \citep{Perez2012, Tazzari2016, Huang2018, Ohashi2019, Sierra2021}, substructures \citep{ALMA2015, Andrews2018, Dong2018}, and planets \citep{Keppler2018}. 
In this work, we intend to go one step further by considering an evolved---yet static---dust population in two-dimensional simulations of smooth protoplanetary disks. 

Our work is motivated by the results of \cite{Dullemond2022}, which show that VSI turbulence in an isothermal disk model is not consistent with observations of thin dust layers in protoplanetary disks. 
In \cite{Pfeil2021}, we have explored the impact of a more realistic cooling time prescription on the strength of VSI turbulence. For this, we assumed the presence of a static, \si{\micro \meter}-sized dust population in the inner parts of a protoplanetary disk (at $\sim \SI{5}{\AU}$). For these setups, we found that the collisional decoupling of the gas and dust particles inhibits thermal relaxation in the disk atmosphere and \rev{thus reduces} VSI turbulence. 
The respective collisional coupling time scale depends on the size distribution and is, thus, sensitive to the fragmentation velocity and other dust properties.
\cite{Fukuhara2021} further studied this effect in models with a more detailed prescription of the dust size distribution. They found that coagulation can indeed inhibit the VSI by depleting the number of small grains that provide radiative cooling.
In their most recent study, \cite{Fukuhara2023} attempted to simulate this in a more self-consistent way, by taking into account the effect of the VSI on the diffusivity and the cooling times.
Since they could not afford to dynamically evolve the dust population within their hydrodynamic simulations, they relied on analytic prescriptions for the cooling time for a static dust size distribution. 

\begin{figure*}
    \centering
    \includegraphics[width=\textwidth]{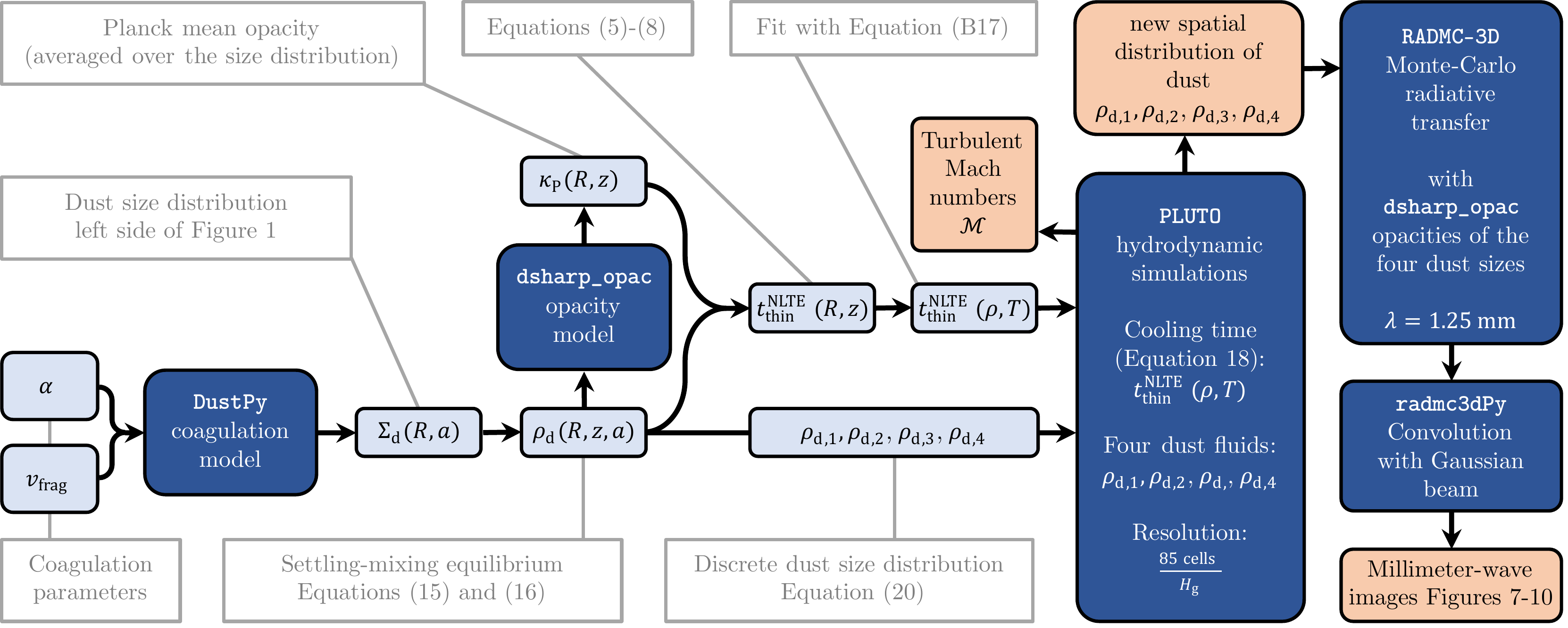}
    \caption{\report{Workflow from our dust coagulation models with \dpy{} to the hydrodynamic simulations with \pluto{} to the radiative transfer modeling with \radmc{}. Used methods and tools are shown as dark blue boxes. Input parameters and intermediate results are shown in light blue. The results of our work are schematically displayed as orange boxes. The details of our methodology are laid out in \autoref{sec:dustpy} (\dpy{} and cooling times), \autoref{sec:pluto} (\pluto{} simulations and results), and \autoref{sec:radmc} (radiative transfer and synthetic observations with \radmc{}).}}
    \label{fig:flowchart}
\end{figure*}

In this work, we study the effect of a more realistic steady-state dust distribution for varying coagulation parameters using \dpy{} \citep{Stammler2022} and \pluto{} \citep{Mignone2007}. We deduce thermal relaxation times from dust coagulation models in \autoref{sec:dustpy} which are then implemented in hydrodynamic simulations, from which we study the VSI activity in \autoref{sec:pluto}.
This makes it possible to study the influence of dust coagulation and the coagulation parameters on VSI turbulence.
\report{These steps are schematically displayed in \autoref{fig:flowchart}.}
In the next step, we introduce passive dust fluids to our simulations in \autoref{sec:dust} to study the effect of the emerging VSI turbulence on the thickness of the dust layer.
To make our results comparable to observations, we create synthetic intensity maps with \radmc{} \citep{Dullemond2012} in \autoref{sec:radmc}.

\section{Theory}
\subsection{Cooling Requirements for the Vertical Shear Instability}
Vertical shear, in the geophysical context also known as thermal wind \citep{Holton2012}, is a consequence of the radial temperature gradient in the vertically stratified protoplanetary disks. 
The temperature gradient itself is maintained by stellar irradiation. 
Consequently, fluid parcels can be displaced upward into a region of lower specific angular momentum experiencing an outward acceleration.
A perturbation along such a trajectory violates Rayleigh's stability criterion and leads to a continued acceleration of the fluid parcel. This mechanism is called the Vertical Shear Instability \citep{Urpin1998} and results in vertically elongated and radially narrow flow patterns.
However, as the gas parcels enter the lower-density regions of the disk atmosphere, they are subjected to buoyancy forces, which, in a stably stratified atmosphere, would lead to an oscillation around the disk midplane. The characteristic frequency of this oscillation is the Brunt-Väisälä frequency 
\begin{equation}
    N_z^2 = -\frac{1}{\rho_\mathrm{g}C_P}\frac{\partial P}{\partial z}\frac{\partial S}{\partial z},
\end{equation}
where $z$ is the distance from the disk midplane, $\rhog$ is the gas density, $P$ is the pressure $S$ is the gas entropy, and $C_P$ is the gas' specific heat capacity at constant pressure.
Thermal relaxation counteracts the restoring force of this oscillation by adjusting a gas parcel's specific entropy to the background.
\hubert{In order for the vertical shear to overcome buoyancy and trigger the VSI, thermal relaxation must be fast. \cite{Lin2015} have shown that vertically global VSI grows the fastest if the cooling timescale fulfills 
\begin{equation}
\label{eq:VSICoolGlobal}
    t_\mathrm{c}< \frac{H_\mathrm{g}}{R}\frac{|\beta_T|}{\gamma-1} \Omega_\mathrm{K}^{-1},
\end{equation}
where $R$ is the distance to the central star, $\beta_T$ is the power law exponent of the temperature profile, $H_\mathrm{g}$ is the pressure scale height, $\Omega_\mathrm{K}$ is the local Keplerian frequency, and $\gamma=\nicefrac{C_P}{C_V}$ is the gas' heat capacity ratio.
\autoref{eq:VSICoolGlobal} was derived under the assumption of a vertically constant thermal relaxation time.
As we specifically consider the height dependence of thermal relaxation, we will use the local definition of a critical cooling time \citep{Urpin2003,Klahr2023} for local VSI modes
\begin{equation}
\label{eq:VSICoolLocal}
    t_\mathrm{c}\lesssim \frac{|r\partial_z\Omega|}{N_z^2}\approx \frac{H_\mathrm{g}}{R}\frac{|\beta_T|\gamma}{2(\gamma-1)}\left(\frac{z}{H_\mathrm{g}}\right)^{-1}\Omega_\mathrm{K}^{-1}.
\end{equation}
In fact, numerical studies like \cite{Manger2021} investigated the dependency of the VSI turbulence on a vertically constant thermal relaxation time and found VSI not to develop for cooling times beyond the critical value for global modes.
This may be due to numerical resolution, as \citet{Lin2015} show that VSI exists for all cooling times, yet at reduced efficiency. \cite{Urpin2003} derived growth rates in this regime, which show a decay proportional to $t_\mathrm{c}^{-1}$. This behavior was recently confirmed in high-resolution\footnote{\texttt{PLUTO-4.2} simulation with $256$ cells per gas scale height, WENO reconstruction, and RK3 time integration \citep{Klahr2023}.} studies of the VSI and other thermal instabilities in disks by \citet{Klahr2023}. It is still subject to investigation how longer growth times will translate into turbulence levels for the non-linear regime, especially in terms of angular momentum transport, diffusion, and gas rms\ velocities. The saturation behavior of VSI and other thermal baroclinic instabilities especially for longer cooling times at sufficient resolution is still being investigated \citep{Latter2018,Cui2022,Klahr2023}.}

\subsection{Optically Thin Thermal Relaxation}
Thermal relaxation of the gas in a protoplanetary disk is mostly achieved via thermal coupling with the dust in a two-stage process. At low temperatures, the emission timescale of the gas molecules is long, which means that cooling \rev{is only possible via} thermal accommodation with the strongly emitting dust particles through collisions. 
\cite{Barranco2018}, derived the thermal relaxation times for the non-LTE case between dust grains and the gas based on the calculation of cooling rates (see \autoref{sec:Appendix_Derivation} for a recap of the derivations). 
For a given dust size distribution $n(a)$, the Sauter mean radius is an instructive parameter in this context, defined as \citep{Sauter1926}
\begin{equation}
    a_\mathrm{S}=\frac{\int n(a) a^3 \, \mathrm{d}a}{\int n(a) a^2 \, \mathrm{d}a}
\end{equation}
where the size integral is executed over the entire size distribution. 
Corresponding to the Sauter mean, we define a respective number density $n_\mathrm{S} =\rhod/\left(\nicefrac{4}{3}\, \pi \rho_\mathrm{m} a_\mathrm{S}^3\right)$ and a collisional cross-section $\sigma_\mathrm{s}=\pi a_\mathrm{S}^2$, where $\rho_\mathrm{m}=\SI{1.67}{\gram \per \cubic \centi \meter}$ is the interior density of the dust grains. With these definitions, we write the thermal accommodation timescale for the gas molecules and the dust grains \citep{Probstein1969, Burke1983} as
\begin{equation}
    t_\mathrm{g}^\mathrm{coll} = \frac{\gamma}{\gamma-1}\frac{1}{n_\mathrm{S}\sigma_\mathrm{S}\Bar{v}_\mathrm{g}},
    \label{eq:tgcoll}
\end{equation}
where $\Bar{v}_\mathrm{g}=c_\mathrm{s}\sqrt{\nicefrac{8}{\pi}}$ is the average gas molecule velocity of a Maxwell--Boltzmann distribution \report{with the isothermal speed of sound $c_\mathrm{s}$}.
Similarly, a timescale for the thermal relaxation of the dust component can be derived, which reads
\begin{equation}
    t_\mathrm{d}^\mathrm{coll} = \left(\frac{\rhod}{\rhog}\right)\left(\frac{C_\mathrm{d}}{\mathrm{C_P}}\right) t_\mathrm{g}^\mathrm{coll},
\end{equation}
with the dust-to-gas density ratio $\nicefrac{\rhod}{\rhog}=\varepsilon$ and the specific heat capacity of the dust particles $C_\mathrm{d}$. As a typical value we pick $C_\mathrm{d}=\SI{800}{\joule\per\kilo\gram\per\kelvin}$, as used by \cite{Barranco2018} \citep[see][]{Wasson1974,Piqueux2021,Biele2022}.
If the collisional coupling is efficient, i.e., temperature perturbations in the gas are transferred to the dust, the thermal equilibrium of the grains will be restored by the emission of radiation. This happens on the black body timescale, depending on the dust density distribution $\rhod(a)$ in units of [\si{\gram \per \cm ^4]} and the respective Planck mean opacity distribution $\planck (a,T)$, in units of [\si{\centi \meter ^2 \per \gram}] 
\begin{equation}
    t_\mathrm{d}^\mathrm{rad} = \frac{\rhod C_\mathrm{d} }{\displaystyle 16 \,\sigma_\mathrm{SB}\, T_\mathrm{eq}^3 } \left(\int \rhod(a)\planck(a,T_\mathrm{eq}) \, \mathrm{d}a\right)^{-1},
\end{equation}
with the Stefan--Boltzmann constant $\sigma_\mathrm{SB}$. 
The total thermal relaxation time of the dust gas mixture can then be calculated following Equation (19) from \cite{Barranco2018}
\begin{equation}
    t_\mathrm{thin}^\mathrm{NLTE} = 2t_{||}\left[1-\sqrt{1-\frac{4t^2_{||}}{t_\mathrm{g}^\mathrm{coll} t_\mathrm{d}^\mathrm{rad}}}\right]^{-1}
\label{eq:tNLTE}
\end{equation}
with $\nicefrac{1}{t_{||}} = \nicefrac{1}{ t_\mathrm{d}^\mathrm{rad}}+\nicefrac{1}{t_\mathrm{d}^\mathrm{coll}}+\nicefrac{1}{t_\mathrm{g}^\mathrm{coll}}$. 
In practice, this means the \rev{slowest channel of energy transfer acts as a bottleneck and the} longest timescale of thermal relaxation determines the cooling time scale of the gas.
\rev{If the dust's emissivity is low, energy cannot be emitted effectively by the grains, and temperature perturbations cannot decay, no matter how well the grains and molecules are coupled ($t_\mathrm{thin}^\mathrm{NLTE}\approx t_\mathrm{d}^\mathrm{rad}$). This situation is unlikely to occur in protoplanetary disks because of the large dust opacities. 
Another case is the collisional decoupling of dust grains and gas molecules. At low densities and in regions where small grains are depleted, heat cannot be transferred between the main carriers of thermal energy (the gas molecules) and the emitters (the dust grains). The high emissivity of the grains does not matter in such a case, since temperature perturbations stay locked in the poorly emitting gas ($t_\mathrm{thin}^\mathrm{NLTE}\approx t_\mathrm{g}^\mathrm{coll}$).}

\citet{Muley2023} introduced a three-temperature radiation transport scheme, which treats dust and gas temperatures separately, yet coupled via collisions. They also find that in most cases the collisional time scale is the most relevant to determine thermal relaxation.

In this case, the cooling time is proportional to the square root of the maximum particle size. 
This can be shown by assuming the size distribution to be a truncated power law with maximum particle size $\amax$, minimum size $\amin$, and power law exponent $p=-3.5$. Then $a_\mathrm{s}=\sqrt{\amax \amin}$ and thus $t_\mathrm{g}^\mathrm{coll}\propto (n_\mathrm{S}\sigma_\mathrm{S})^{-1} \propto \sqrt{\amax}$. Sticking collisions between grains typically increase the maximum particle size until a fragmentation--coagulation equilibrium is reached. In this case, $\amax\approx a_\mathrm{frac}\propto v_\mathrm{frag}^{2}$ holds \citep{Birnstiel2012}, and we deduce that the collisional timescale \report{is directly proportional to the} fragmentation velocity in this case.
Laboratory experiments aim to determine the actual value of $v_\mathrm{frag}$ which is dependent on the composition and porosity of grains \citep{Blum2000, Wurm2001, Blum2006, Musiolik2019}. Typical values lie within a range of \SIrange{100}{1000}{\centi\meter \per \second}.

An additional uncertainty arises from the unknown relative grain velocities, which depend on the strength of turbulence, differential drift, and settling. Especially the strength of turbulence in protoplanetary disks is highly uncertain and also a subject of this article. The simplest assumption for the turbulent transfer of energy across length scales is the Kolmogorov cascade. For the resulting energy spectrum, relative grain velocities can be approximated as $\delta v\approx \sqrt{3\alpha \St}c_\mathrm{s}$ \citep{Ormel2007}, \report{with the Stokes number $\St$ (see \autoref{eq:Stokes}). This} is the underlying assumption for the derivation of $a_\mathrm{frag}$. 
\rev{
In this turbulence prescription, which is Based on the assumption of a mixing length model \citep{Prandtl1925}, turbulent stresses result in an effective viscosity 
\begin{equation}
    \nu =\alpha c_\mathrm{s} H,
\end{equation}
where $c_\mathrm{s}$ is the local sound speed and $H$ is the pressure scale height of the disk \citep{Shakura1973}.
From this, turbulent rms\ velocities can be related to $\alpha$ 
by assuming a turbulent correlation time of $\Omega_\mathrm{K}^{-1}$ via
%
\begin{equation}
\label{eq:alpha}
    \alpha = \frac{\langle v_\mathrm{turb} ^2 \rangle}{c_\mathrm{s}^2}.
\end{equation}
}
With this, $a_\mathrm{frag}\propto \alpha^{-1}$, implying $t_\mathrm{g}^\mathrm{coll}\propto \alpha^{-\nicefrac{1}{2}}$. 
Low $\alpha$ therefore corresponds to longer cooling times, as a consequence of the presence of larger particles. Additionally, lower levels of turbulence correspond to smaller dust scale heights, leading to a depletion of the upper layers and an additional dampening of the VSI in these regions.

\cite{Fukuhara2021} investigated the effect of varying maximum particle sizes throughout a protoplanetary disk and found that the presence of VSI depends on particle sizes via the cooling time dependency.  

In the following sections, we investigate this effect through the use of more realistic dust coagulation models and subsequent hydrodynamic simulations. 
We aim to determine the implications for the interpretation of observational data and the respective feedback onto the dust layer by turbulent mixing through the VSI. 

\section[DustPy Coagulation Models]{\dpy{} Coagulation Models}
\label{sec:dustpy}
In the previous sections, we discussed the importance of thermal relaxation for the VSI. We have also highlighted that the cooling times are highly sensitive to the present dust population, most importantly, the maximum particle size.

In this section, we present a series of dust coagulation simulations, conducted with \dpy{}, that further illustrate the impact of dust coagulation on the cooling times. 
We use the output of these simulations to calculate cooling time distributions for our subsequent hydrodynamic simulations with the \pluto{} code.

For our disk model we employ the standard \cite{LyndenBell1974} profile for a solar-mass star and a \SI{0.05}{\Msol} disk \report{with dust-to-gas ratio (metallicity) $\mathcal{Z}=0.01$} (see \autoref{tab:StellarParams})
\begin{equation}
\Sigma_\mathrm{g} =\frac{M_{\text{d}}(1+\beta_{\Sigma})}{2\pi R_\mathrm{c}^2} \left(\frac{R}{R_\mathrm{c}}\right)^{\beta_{\Sigma}}\exp{\left[-\left(\frac{R}{R_\mathrm{c}}\right)^{2+\beta_{\Sigma}}\right]}.
\end{equation}
We set the radial column density gradient to $\beta_\Sigma=-0.85$, and the cutoff radius to $R_\mathrm{c}=\SI{100}{\AU}$. 
Our radial temperature profile is determined by passive stellar irradiation and assumed to be constant in the vertical direction \citep[see][]{Chiang1997, DAlessio1998, Dullemond2018}
\begin{equation}
    T= \left(\frac{ \varphi L_*}{4\pi  R^2 \sigma_\mathrm{SB}}\right)^{\nicefrac{1}{4}},
    \label{eq:Irrad}
\end{equation}
where $L_*$ is stellar luminosity, and $\varphi=0.02$ is the flaring angle. 
Gas evolution and dust drift alter the dust size distribution in protoplanetary disks. The overall effect of these transport phenomena on the shape of the distribution is, however, most relevant in the final stages of disk evolution, when the growth front has reached the outer disk edge and the mass budget is quickly decreasing \citep[i.e., when the dust accretion rate is no longer radially constant,][]{Birnstiel2014}. At what point in time after disk formation this becomes relevant is dependent on the disk's size, its radial structure, the dust-to-gas ratio, the strength of turbulence, the fragmentation velocity, etc.
In this study, we are interested in the effect of dust coagulation on the cooling times and, through the cooling times, on the VSI. In the inner parts of the disk, a steady state distribution, determined by fragmentation and coagulation, will be reached and approximately maintained as long as the outer disk edge is not yet moving inward. 
We have therefore decided to completely disregard any transport effects (except the vertical settling--mixing equilibrium).  
We are thus calculating a steady state dust distribution for each parameter set that is only determined by fragmentation and coagulation. The output of our models is, therefore, time-independent once the equilibrium size distribution is reached at each radius. In that way, we avoid selecting an arbitrary simulation snapshot.

\report{Note that this is still an idealized assumption. In reality, radial drift and gas evolution could slightly alter the radial structure and the size distributions at similar timescales. Typically, drift-limited size distributions are slightly steeper than in the fragmentation limit \citep{Birnstiel2011}. In recent studies, the VSI itself was also shown to alter the radial disk structure \citep{Manger2021}.}

Our \dpy{} models are run for \SI{e5}{\years}, after which coagulation--fragmentation equilibrium is reached at every radial grid cell.

We conduct simulations for three different fragmentation velocities $v_\mathrm{frag}=\SIlist[list-units = single]{100;200;400}{\centi \meter \per \second}$ and for a turbulence parameter $\alpha =10^{-3}$.
Additionally we probe two different turbulent \report{diffusivities} with $\alpha=\SIlist{e-4;e-2}{}$, at $v_\mathrm{frag}=\SI{100}{\centi \meter \per \second}$. 
\report{At this point we do not further specify the origin of the diffusivity $\alpha$, making it a free parameter for the coagulation models.}
We show the resulting dust size distribution at \SI{50}{\AU} and \SI{100}{\AU} on the left-hand-side of \autoref{fig:CoolingTimes} \rev{and some key particle properties are shown in \autoref{tab:StellarParams}}.
\report{We can see that the particles grow to larger sizes at smaller distances to the central star, in accordance with analytic estimates of the fragmentation-limited particle size \citep{Birnstiel2012}. The respective size distributions can be approximated with power laws with exponents $-p\approx\SIrange{3.6}{3.7}{}$. These values lie within the typical range for fragmentation-limited size distributions derived by \cite{Birnstiel2011}.}

\subsection{Thermal Relaxation Times Derived from Dust Coagulation Simulations}
We derive the vertical structure from these, vertically integrated, \dpy{} models by assuming vertical hydrostatic equilibrium for the gas and vertical settling--mixing equilibrium for the dust. 
Gas densities thus follow
\begin{equation}
    \rho = \rho_\mathrm{mid} \exp{\left[\left(\frac{H_\mathrm{g}}{R}\right)^{-2}\left(\frac{R}{\sqrt{R^2+z^2}} - 1\right)\right]},
\end{equation}
with gas scale height $H_\mathrm{g}$ and $\rho_\mathrm{mid}=\nicefrac{\Sigma(R)}{\sqrt{2\pi H_\mathrm{g}^2}}$. We assume an ideal equation of state $P=\rho c_\mathrm{s}^2$.
The vertical dust distribution is determined by the diffusion parameter $\delta$ and the Stokes number of the individual size bins on the size distribution, which is defined as
\begin{equation}
    \St = \frac{\pi}{2}\frac{a\rho_\mathrm{m}}{\Sigma_\mathrm{g}}.
    \label{eq:Stokes}
\end{equation}
Volume dust densities for each size are then derived by calculating the dust scale height
\begin{align}
    H_\mathrm{d} &= H_\mathrm{g} \sqrt{\frac{\delta}{\delta+\St}} \\
    \rho_\mathrm{d} &= \rho_\mathrm{d,mid} \exp{\left[\left(\frac{H_\mathrm{d}}{R}\right)^{-2}\left(\frac{R}{\sqrt{R^2+z^2}} - 1\right)\right]}, \label{eq:rhodust}
\end{align}
with $\rho_\mathrm{d,mid}=\nicefrac{\Sigma_\mathrm{d}(R)}{\sqrt{2\pi H_\mathrm{d}^2}}$. 

\begin{figure*}
    \centering
    \includegraphics{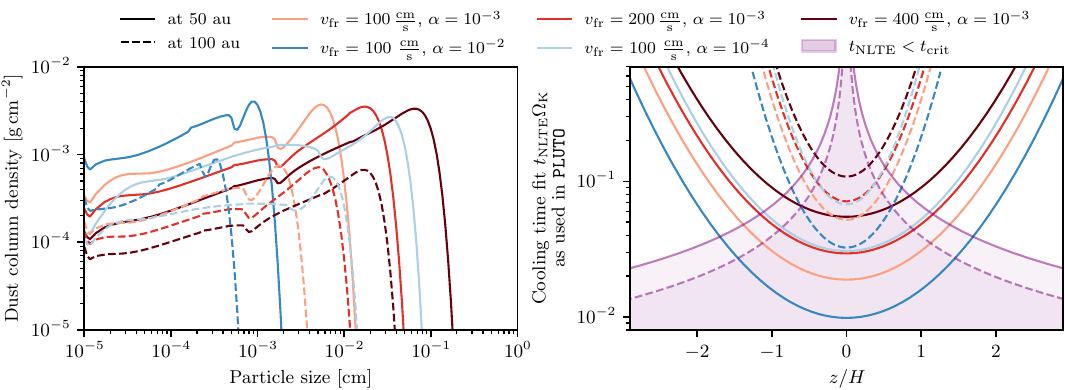}
    \caption{Dust size distributions at \SI{50}{\AU} (solid lines) and \SI{100}{\AU} (dashed lines) of our \dpy{} models (left side). On the right-hand side, we show the respective vertical cooling time profiles, assuming vertical settling--mixing equilibrium for the given $\alpha$ and the critical VSI cooling time. Models with larger particles also exhibit longer cooling times due to collisional decoupling between dust and gas. We also show the height-dependent cooling time for local VSI modes as purple lines (see \autoref{eq:VSICoolLocal}).}
    \label{fig:CoolingTimes}
\end{figure*}

The resulting temperature and density structure is used to calculate the Planck mean opacities of the dust. We use the DSHARP opacity model by \cite{Birnstiel2018} as implemented in the \texttt{dsharp\_opac} python package \rev{with the standard DSHARP particle properties}.  
Thermal relaxation times of the gas can then be calculated from the disk structure and opacities via equations (\ref{eq:tgcoll})--(\ref{eq:tNLTE}).
For the given parameters in our simulations, we find that the thermal relaxation time is limited by the collision timescale outside of $\sim\SI{10}{\AU}$. At smaller radii, the disk might become optically thick, meaning the relaxation time of temperature perturbations depends on \rev{the respective length scale}. We are therefore only modeling the parts of the disk around \SI{50}{\AU}, where thermal relaxation operates in the optically thin regime.
\autoref{fig:CoolingTimes} shows the size distributions and the vertical profile of the thermal relaxation times for the respective coagulation and turbulence parameters at \SI{50}{\AU} and \SI{100}{\AU}. We find that the cooling times increase with height above the midplane. The reason for this is that cooling is achieved via collisions between dust particles and gas molecules, which become rarer at lower densities. 
This also means that models with larger particles have longer thermal relaxation times because of the reduced number densities of dust particles and the stronger settling. 
Higher fragmentation velocities are counteracting the VSI.
Likewise, models with weaker turbulence parameter $\alpha$ can also be expected to have less VSI activity, as demonstrated by our numerical simulations.

\section[PLUTO Simulations based on Coagulation Model]{\pluto\ Simulations based on Coagulation Models}
\label{sec:pluto}

We set up axisymmetric \pluto{} simulations with the same radial structure as our \dpy{} models to study the evolution of VSI with the respective model's cooling times.
 Pressure forces act in the outward direction of the disk and therefore decrease the equilibrium rotation frequency of the gas, especially at the steep outer edge of the disk. We define our hydrostatic initial rotation profile accordingly as
\begin{align}
\frac{\Omega^2(R,z)}{\Omega_\mathrm{K}^2} = &\left(\frac{H_\mathrm{g}}{R}\right)^2\left(\beta_T+\beta_\rho-(\beta_\Sigma + 2)\left(\frac{R}{R_\mathrm{c}}\right)^{\beta_\Sigma+2}\right) \nonumber \\ 
&- \frac{\beta_T R}{\sqrt{R^2+z^2}} + \beta_T + 1 ,
\end{align}
\report{where $\beta_\rho$ is the power law exponent of the midplane gas density $\rho\propto R^{\beta_\rho}$ and $\beta_{T}$ is the power law exponent of the radial temperature profile $T\propto R^{\beta_T}$}.
Thermal relaxation is realized as in \cite{Pfeil2021}, by analytically relaxing the gas pressure toward the equilibrium profile (determined by stellar irradiation). Density is kept constant in this cooling step, which makes a relaxation in pressure equal to a relaxation in temperature for an ideal equation of state. 
\begin{align}
    P^{(n+1)} &= P_\mathrm{eq} + (P^{(n)}-P_\mathrm{eq})\exp\left(-\frac{\Delta t}{t_\mathrm{thin}^{\text{NLTE}}}\right) \nonumber \\
   \overset{            \begin{subarray}{c}
                \mathrm{const.}\\
                \rho
            \end{subarray}}{\Longleftrightarrow} T^{(n+1)} &= T_\mathrm{eq} + (T^{(n)}-T_\mathrm{eq})\exp\left(-\frac{\Delta t}{t_\mathrm{thin}^{\text{NLTE}}}\right),
\end{align}
where $(n)$ denotes the number of the current simulation timestep of length $\Delta t$.
The equilibrium temperature $T_\mathrm{eq}$ is defined by stellar irradiation (\autoref{eq:Irrad}).
Cooling times, presented in the previous section, are derived from \dpy{} simulations (see \autoref{fig:CoolingTimes}) and subsequently fitted as a function of local gas density and temperature for each simulation (for a detailed description of the fits, see \autoref{sec:Appendix_Fits}). 

\report{Fitting the spatial distributions of the thermal relaxation times as functions of density and temperature also introduces uncertainties in the cooling times for PLUTO. For all models except one, these errors lie within $\SI{25}{\percent}$ with respect to the real distribution of cooling times. For the case of the most settled particles ($v_\mathrm{frag}=\SI{100}{\centi \meter \per \second}$, $\alpha=10^{-4}$), however, the fitting function seems to diverge further from the real distribution and the fit deviates up to \SI{58}{\percent} from the cooling times close to the midplane. This is likely due to the difference between this particular highly settled model and the other less settled cases. 
Since the cooling times vary over several orders of magnitude throughout the simulation domain and between the models, we deem this uncertainty acceptable--also because the overall distribution of cooling times is still well reproduced (this can be seen in the matching contours in \autoref{fig:CoolingFits}). It is worth noting, however, that in this work, we only study the overall trends of VSI turbulence with the coagulation parameters and do not aim to exactly reproduce specific systems or observations.}

The resulting analytic \report{cooling time} prescriptions are used within our \pluto{} simulations to calculate $t_\mathrm{thin}^{\text{NLTE}}$ from the local disk structure. Since cooling is dominated by the small grains, which predominantly move along with the gas, minor disturbances in the gas densities, as caused by the VSI, can also influence the cooling times in this model. We emphasize that this is a minor effect in our simulation, and does not have an impact on the resulting turbulence.
\report{It should be noted}, that our cooling time prescription, which is derived from dust coagulation models, is static throughout the simulation.

\report{Although our coagulation models assumed a certain turbulent diffusivity delta to calculate relative particle velocities, we set up our hydrodynamic simulations to be inviscid. This is because we want to study the onset of the VSI and the resulting turbulence strength. Applying the same diffusivities as for the coagulation models ($\delta=\SIrange{1e-4}{1e-2}{}$) as viscosity in \pluto{} would likely stop the VSI from emerging in the first place \citep{Barker2015}. Note that setting up viscous simulations would also not be fully self-consistent since relative particle velocities in \dpy{} are inferred from perfectly isotropic turbulence and the resulting Kolmogorov cascade, which would not be the case for the developing VSI turbulence in our simulations.}

We carry out the calculations for 500 orbital periods at \SI{50}{\AU} ($=\SI{176777}{\years}$). Simulation domains are set up in spherical coordinates and extend from \SIrange[range-units = single]{25}{150}{\AU} in the radial direction, and over $\pm 3$ pressure scale heights from the midplane of the disk in the polar direction. We resolve one scale height at \SI{50}{\AU} with 85 cells and employ logarithmic griding in the radial direction to preserve the cells' aspect ratios, resulting in a $2011_r \times 513_\vartheta$ grid.
Periodic boundary conditions are set up in the azimuthal direction with only one grid cell, making our simulations axisymmetric. Radial and polar boundaries are setup up as reflective for the orthogonal velocity components and as zero-gradient for the respective tangential velocity components.
Pressure and density in the boundary cells are kept at the initial condition.

\begin{figure*}[ht]
    \includegraphics{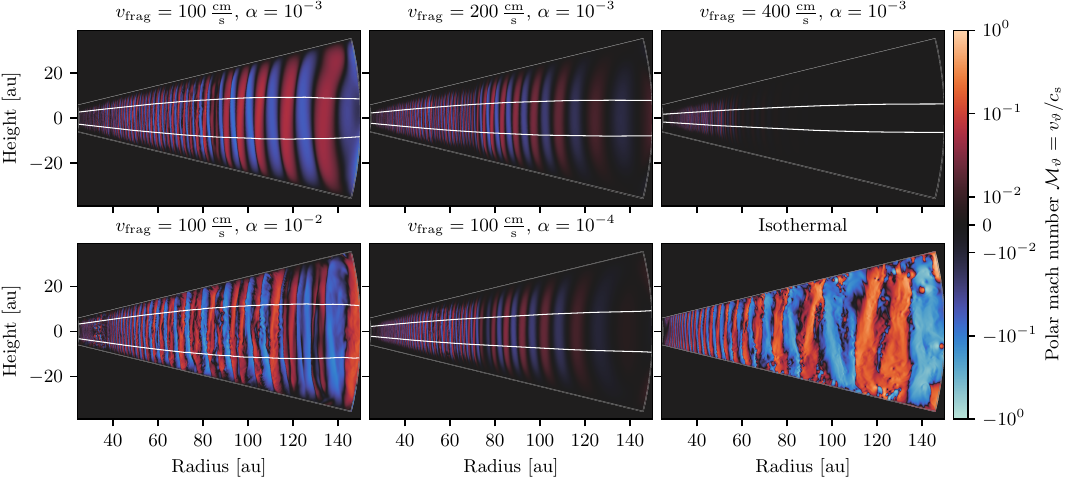}
    \caption{Vertical velocities in units of the local speed of sound in our six \pluto{} runs after 500 orbital time scales at \SI{50}{\AU}. The isothermal run shows a snapshot after only 200 orbits. White contours mark the position at which the critical cooling time for the VSI is reached (\autoref{eq:VSICoolLocal}), i.e., VSI is theoretically possible within the white lobes.}
    \label{fig:VSIVel}
\end{figure*}

In \autoref{fig:VSIVel}, we show the vertical velocities in our simulations at the end of the simulation time. It is evident, that the spatially varying cooling times set constraints on where the VSI can be active and where vertical motions are suppressed by buoyancy. As a comparison, we also show an isothermal simulation (i.e., ideal VSI), in which the resulting turbulence is present in the entire simulation domain and at higher turbulent Mach numbers.
For the case of $v_\mathrm{fr}=\SI{400}{\centi \meter \per \second}$ and $\alpha=10^{-3}$, we find the disk to be completely quiescent outside of $\sim \SI{80}{\AU}$, due to the long cooling times. In this case, dust would settle into a very thin layer in the outer disk, which we will further investigate in the next sections.
Similarly, the disk regions outside of $\sim \SI{100}{\AU}$ show only very little VSI activity for the coagulation model with $v_\mathrm{fr}=\SI{100}{\centi \meter \per \second}$ and $\alpha=10^{-4}$.

\begin{figure*}[ht]
    \includegraphics[width=\textwidth]{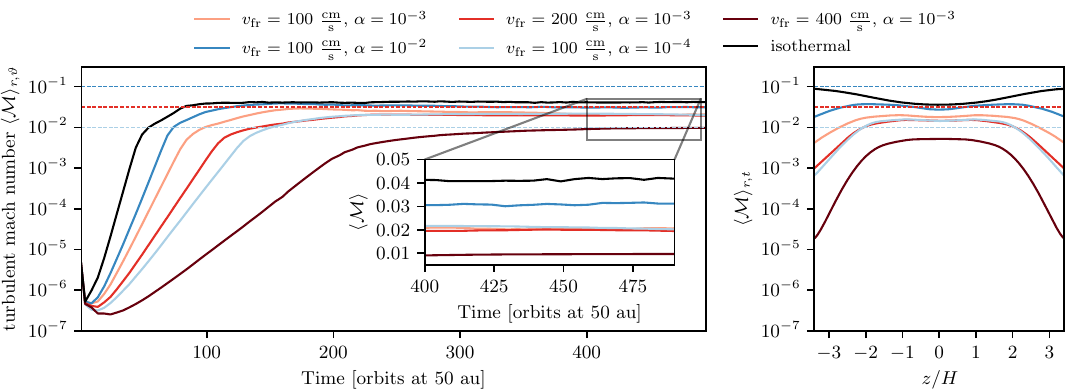}
    \caption{Time evolution of vertical shear instability simulations based on the different dust models. Turbulent Mach numbers are shown as a function of time (radially and vertically Favre-averaged) and as a function of height above the midplane (time-averaged and radially Favre-averaged). In models with larger particles, cooling times are generally longer, which results in lower growth rates and lower Mach number turbulence. \rev{The vertical profiles on the right-hand side change accordingly. Cooling times in models with larger maximum particle size \report{increase} more rapidly with height above the midplane, which also cuts off the VSI turbulence. Isothermal models typically have vertically increasing turbulent velocities.} \hubert{The three dashed horizontal lines show the Mach numbers corresponding to the three $\alpha$ values that we assumed for our coagulation models (see \autoref{eq:alpha}). Note that \report{the conversion between turbulent Mach numbers and diffusivities} assumes a perfect Kolmogorov turbulence spectrum (see discussion in \autoref{sec:discussion}), \report{which is likely not given for the anisotropic VSI turbulence}.}}
    \label{fig:MachTimeVertical}
\end{figure*}

To characterize the development and strength of the VSI turbulence, we measure the Favre-averaged (i.e. density-weighted) turbulent Mach numbers \report{over the whole} simulation domains, \report{where the average in a direction $x$ (polar, radial, or both) is defined as
\begin{equation}
    \langle \mathcal{M}\rangle_x = \frac{\int \frac{\sqrt{v_r^2+v_\vartheta^2}}{c_\mathrm{s}} \rho\,\mathrm{d}x}{\int \rho \,\mathrm{d}x},
\end{equation}
\reporttwo{where $v_r$ and $v_\vartheta$ represent the radial and polar velocity components. Since our simulations are set up hydrostatically, these components measure turbulent fluctuations caused by the VSI.}
}
While velocities in our isothermal simulation saturate after $\sim$ 100 orbits at $\langle\mathcal{M}\rangle \approx 4\times 10^{-2}$, all other, non-ideal simulations, reach lower Mach numbers and have longer growth time scales (see \autoref{fig:MachTimeVertical}). 
The vertical profile of the Mach numbers shows the typical vertical increase and a sharp upper cutoff, similar to the results in \cite{Pfeil2021}. The collisional decoupling of dust particles and gas molecules is the reason for this behavior. 
\report{\autoref{fig:MachTimeVertical} also shows the three Mach numbers corresponding to the diffusivities chosen to calculate turbulent relative velocities between particles in our coagulation model ($\alpha$= \SIlist{1e-4;1e-3;1e-2}{}). As can be seen, the three lines do not exactly correspond to the measured Mach numbers of our simulations. This is, however, also not to be expected, since the direct conversion between Mach numbers and particle collision speed (see \autoref{eq:alpha}) assumes a perfect Kolmogorov spectrum and, thus, isotropic turbulence which is not given for the VSI. The calculation of collision speeds would furthermore depend on the correlation time spectrum which was not taken into account here.}

\begin{figure}
    \centering
    \includegraphics{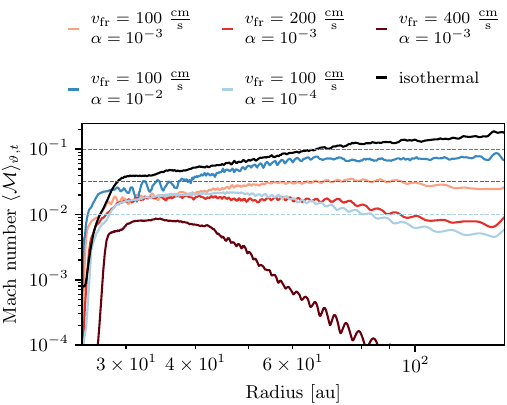}
    \caption{Radial dependency of the turbulent Mach numbers in a polar and time Favre average over 200 orbits in our VSi simulations. VSI simulations based on \dpy{} models with larger particles have lower levels of turbulence. For our model with the largest particles $v_\mathrm{fr}=\SI{400}{\centi \meter \per \second}$ and $\alpha=10^{-3}$, the outer disk, beyond \SI{80}{\AU} is completely quiescent. \hubert{The three dashed horizontal lines show the Mach numbers corresponding to the three $\alpha$ values that we assumed for our coagulation models (see \autoref{eq:alpha}). Note that such a conversion assumes a perfect Kolmogorov turbulence spectrum (see discussion in \autoref{sec:discussion}).}}
    \label{fig:MachRadial}
\end{figure}

\autoref{fig:MachRadial} depicts the radial dependence of the Mach numbers in our simulations. 
The lowest turbulence levels of $\langle\mathcal{M}\rangle \approx 8\times 10^{-3}$ are reached in our simulations based on the \dpy{} model with $\alpha=10^{-3}$ and $v_\mathrm{fr}=\SI{400}{\centi \meter \per \second}$, i.e., in the model with the largest particles ($\amax(\SI{50}{\AU})\approx \SI{0.14}{\centi \meter}$).
For this simulation, we observe a decrease in turbulence outside of \SI{40}{\AU}. At $\SI{60}{\AU}$, turbulent Mach numbers have already decreased by a factor 10 compared to the inner regions.
Also our models with $v_\mathrm{fr}=\SI{200}{\centi \meter \per \second}$ and $\alpha=10^{-3}$ and the model with $v_\mathrm{fr}=\SI{100}{\centi \meter \per \second}$ and $\alpha=10^{-4}$ show a radially decreasing level of turbulence in the outer disk.

We conclude that the level of VSI turbulence is highly dependent on the physical details of the dust coagulation process. 
If dust grains can grow up to the fragmentation limit---which is to be expected in most parts of protoplanetary disks \rev{in the early evolutionary stages}---we can expect weak collisional coupling between dust grains and gas molecules in the optically thin, outer regions, leading to inefficient cooling and only weak VSI turbulence. 
The magnitude of the impact of dust coagulation on the hydrodynamic turbulence depends mostly on the maximum size of the grains, where larger grains correspond to less cooling and, thus, stronger damping of VSI.

\subsection[Dust Dynamics in the PLUTO Simulations]{Dust Dynamics in the \pluto{} Simulations}
\label{sec:dust}
In the previous section, we have shown that the VSI activity in protoplanetary disks is highly sensitive to the properties of the present dust grain population, especially the largest grain size. 
However, we cannot directly infer the VSI's feedback on the dust population. \cite{Dullemond2022} have clearly shown that the ideal VSI is inconsistent with the observed thickness of protoplanetary disks in millimeter-wave observations with ALMA \citep{Villenave2020, Villenave2022}.
Our simulations show that the level of turbulent vertical velocities can vary by orders of magnitude across the disk, depending on the details of the dust size distribution.

In this section, we explore how these different levels of turbulence impact the thickness of the dust layer.
For this, we restart the simulations after the VSI has reached a saturated level of turbulence. We add four dust fluids, resembling a power law size distribution \report{$n(a)\propto a^p$, and thus $\Sigma_\mathrm{d}(a)\propto n(a)m(a) \propto a^{p+3}$. Normalizing to the total dust column density (column dust-to-gas ratio $\mathcal{Z}=0.01$) and integrating the distribution over the size bin $i$ with boundaries $a_i$ and $a_{i+1}$, we get
\begin{align}
\begin{split}
        \Sigma_{\mathrm{d},i} =  
        \begin{cases}
    \Sigma_{\mathrm{d,tot}}\frac{a_{i+1}^{p+4}-a_i^{p+4}}{\amax^{p+4}-\amin^{p+4}} & \text{for } p\neq -4 \\
    \Sigma_{\mathrm{d,tot}}\frac{\log(a_{i+1})-\log(a_i)}{\log(\amax)-\log(\amin)} & \text{for } p=-4
  \end{cases} .   \label{eq:distr}
\end{split}
\end{align}
}
The maximum grain sizes $\amax$ and exponents $p$ are derived from the underlying \dpy{} models (\rev{measured at a distance of \SI{50}{\AU} as the size including \SI{99.9}{\percent} of the dust mass}, see \autoref{tab:sizes}). 
\report{Similar to the \dpy{} simulations}, the minimum grain size is set to \SI{0.1}{\micro \meter}, \report{which is a typical size assumed for monomers in protoplanetary disks \citep{Tazaki2022} and which is constant throughout the simulations}.
We divide the power law size distribution into four sections, equally spaced in logarithmic size space \report{between $\amin$ and $\amax$}. 
The initial vertical dust distribution is determined by the midplane Stokes numbers and the level of turbulence assumed in the respective \dpy{} runs, \report{follwing \autoref{eq:rhodust}}. Dust is allowed to flow in from the outer boundary of the simulation domain with the initial vertical distribution.
As to the time of this work, the \pluto{} code has no built-in dust fluids. Therefore, we make use of the available gas tracer fluids. To model radial dust drift and vertical settling we modify the tracer fluxes according to the respective grain sizes' relative velocity to the gas, which is given by the prescriptions of \cite{Nakagawa1986} \rev{(terminal velocity approximation)}. Each dust fluid is advected with the \rev{gas velocity plus the drift correction of the mass-averaged size of the respective size bin}. In \autoref{sec:Appendix_Tests} we present tests of this method that verify its accuracy. 

\begin{deluxetable*}{ccccccccccccc}
\tablecaption{Dust coagulation parameters of our five \dpy{} simulations and the respective maximum particle size measured at \SI{50}{\AU} in the \dpy{} simulation.  \label{tab:sizes} \label{tab:StellarParams}}
\tablehead{\colhead{$M_*$} & \colhead{$R_*$} & \colhead{$T_*$} & \colhead{$M_\mathrm{disk,g}$} & \colhead{$\mathcal{Z}$} & \colhead{$v_\mathrm{fr}$} & \colhead{$\alpha_\mathrm{turb}$} & \colhead{$\rho_\mathrm{m}$} & \colhead{$\amin$} & \colhead{$\amax$ (\SI{50}{\AU})} & \colhead{$\St_\mathrm{max}$ (\SI{50}{\AU})} & \colhead{$a_\mathrm{s}$ (\SI{50}{\AU})} & \colhead{$\St_\mathrm{s}$ (\SI{50}{\AU})} \\ [-0.2cm]
\colhead{[$M_\odot$]} & \colhead{[$R_\odot$]} & \colhead{[\si{\kelvin}]} & \colhead{[$M_*$] } &  & \colhead{[\si{\centi \meter \per \second}]} &  & \colhead{$\mathrm[\si{\gram \per\cubic\centi\meter}]$} & [\si{\centi \meter}] & \colhead{[\si{\centi \meter}]} &  & \colhead{[\si{\centi \meter}]} & }
\startdata
         1 & 2 & 5772 & 0.05 & 0.01 & 100 & $10^{-3}$ & 1.67 & $10^{-5}$ &\expnum{ 0.0107977516232771 } & \expnum{ 0.0025113868507170675 } & \expnum{ 0.00019916111110875457 } & \expnum{ 4.632173558562798e-05 } \\ 
         " & " & " & " & " & 200 & $10^{-3}$ & " & " &\expnum{ 0.03981071705534976 } & \expnum{ 0.009259345354350655 } & \expnum{ 0.0004295029706271335 } & \expnum{ 9.989562183034675e-05 } \\
         " & " & " & " & " & 400 & $10^{-3}$ & " & " &\expnum{ 0.146779926762207 } & \expnum{ 0.03413869765490539 } & \expnum{ 0.0012600382024171158 } & \expnum{ 0.0002930650271793446 } \\
         " & " & " & " & " & 100 & $10^{-4}$ & " & " &\expnum{ 0.06309573444801937 } & \expnum{ 0.014675073418756788 } & \expnum{ 0.0009685166350693979 } & \expnum{ 0.00022526170511003303 } \\ 
         " & " & " & " & " & 100 & $10^{-2}$ & " & " &\expnum{ 0.001584893192461115 } & \expnum{ 0.0003686211780198209 } & \expnum{ 8.519752028663313e-05 } & \expnum{ 1.9815600471889103e-05 } \\
\enddata
\vspace{1ex}
{\footnotesize \textbf{Note}. This value was used as a maximum particle size in our \pluto{} simulations with dust. We also show the respective Sauter mean radii and Stokes numbers.}
\end{deluxetable*}

\begin{figure*}
    \includegraphics{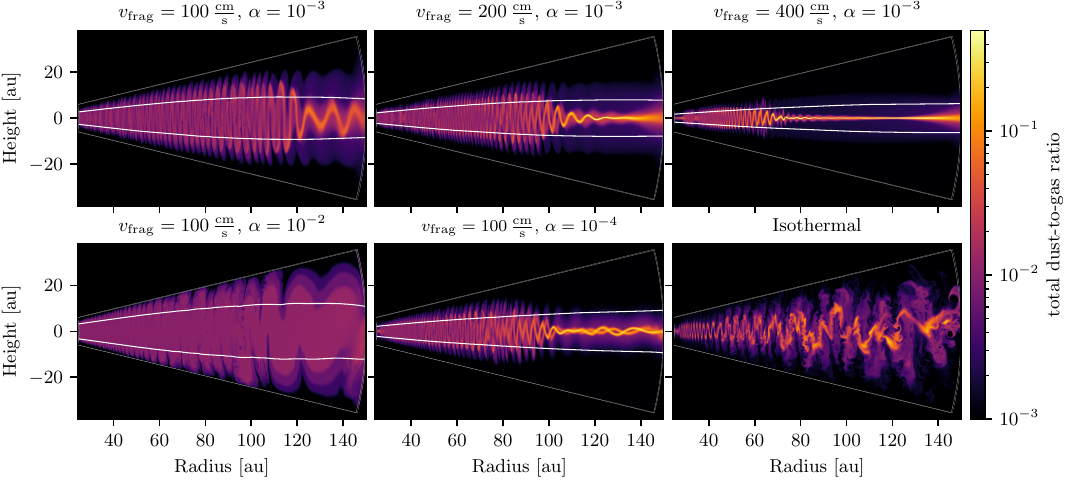}
    \caption{Total dust to gas ratios in VSI simulations restarted after 425 orbits with four passive dust fluids. Each simulation is started with a dust distribution similar to the one derived from the respective \dpy{} simulations. Snapshots are taken after 150 orbits of evolution. White contours mark the position at which the critical cooling time for the VSI is reached (\autoref{eq:VSICoolLocal}), i.e., VSI is theoretically possible within the white lobes.}
    \label{fig:VSIDust}
\end{figure*}

We \rev{continue the previous, gas-only,} VSI simulations with dust for \rev{another} 150 orbits (measured at \SI{50}{\AU}).
\autoref{fig:VSIDust} depicts the distribution of dust-to-gas ratios in our simulations after 150 orbits. In our model with $\alpha=10^{-3}$ and $v_\mathrm{fr}=\SI{400}{\centi \meter \per \second}$, we have the largest particles of $\approx \SI{0.14}{\centi \meter}$ radius, while the smallest particles are present in the model with $\alpha=10^{-2}$ and $v_\mathrm{fr}=\SI{100}{\centi \meter \per \second}$, with a maximum size of $\approx \SI{15}{\micro \meter}$ (see \autoref{tab:sizes}).
As a comparison, we initialize the isothermal simulation with the largest grains, to get an estimate of the effect of ideal VSI on a grown dust population \citep[as in][]{Dullemond2022}.
The effect of the different levels of VSI turbulence, depending on the coagulation parameters and the respective thermal relaxation times becomes visible in the dust-to-gas ratios, where the simulations with larger particles, longer cooling times, and less VSI turbulence have more settled dust layers. Especially the outer disk regions are affected by this, as can be seen in the cases with $v_\mathrm{fr}>\SI{100}{\centi \meter\per \second}$ and $\alpha<10^{-2}$.

\report{We can furthermore see, that the isothermal simulation provides a good approximation for the models with the smallest particles. This is to be expected because the models with the smallest particles also have the shortest cooling times, making the VSI modes almost isothermal.}

\begin{figure*}
    \centering
    \includegraphics{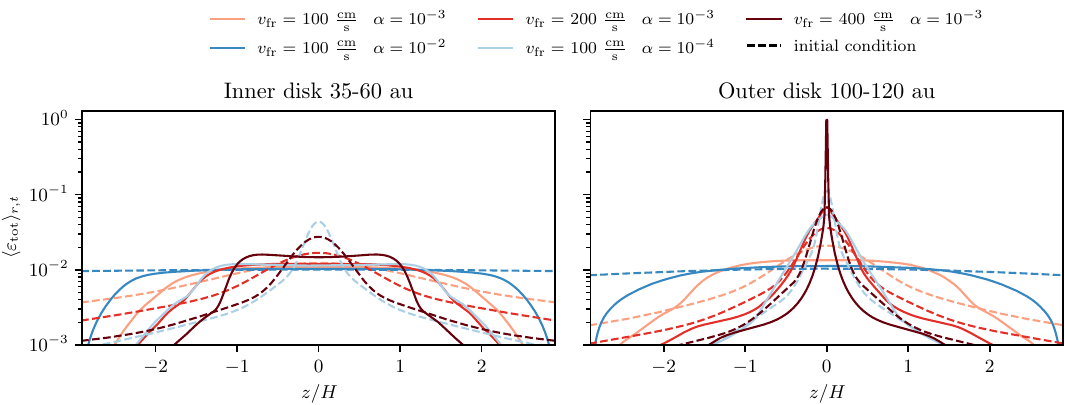}
    \caption{Radially and time-averaged dust-to-gas ratios in the inner and outer parts of our simulations. The inner regions are VSI active, forming plateau-like dust distributions in all simulations with a cutoff at the edges of the VSI active zones. The outer disk regions appear much more settled in that cases of $v_\mathrm{fr}=\SI{200}{\centi\meter\per\second}$ and $v_\mathrm{fr}=\SI{400}{\centi\meter\per\second}$, in which the outer regions are quiescent.}
    \label{fig:dtgratios}
\end{figure*}

To visualize the clear distinction between the inner VSI active region and the outer VSI inactive regions, we plot the time and radially averaged total dust-to-gas ratios in \autoref{fig:dtgratios}.
For the models with fully VSI active disks, we find flat top, or double-peaked dust distributions throughout the entire disks.
In contrast, models with larger grains and inactive outer disks, show flat-topped, or double-peaked profiles in the inner disk regions and highly settled outer regions.

\hubert{A perfect flat-top distribution would indicate spatially homogeneous diffusion and could easily be fitted by an analytic expression  \citep[see \autoref{eq:eps_vert}, ][]{Fromang2009}. The double hump, on the other hand, cannot be a feature of iso-tropic turbulence and reflects the action of the quasi-periodic VSI motions.}

At this point, we can only speculate what the feedback of these dust distributions onto the VSI would be. \report{\cite{Lin2017} and \cite{Lin2019} studied the influence that dust back-reaction could have on the VSI and found that this process generally damps the VSI turbulence.} For the highly settled cases, with midplane dust-to-gas ratios near unity, one would have to include hydrodynamic back-reaction, as in the work by \citet{Schafer2020, Schafer2022}. In these scenarios, the presence of VSI would probably be further inhibited by the hydrodynamic feedback of the dust onto the gas.
Cooling times would also increase significantly in these regions. The areas above the midplane would be in the collision-limited regime, whereas the midplane could become optically thick (see \autoref{sec:radmc}).

\section{Radiative Transfer Post Processing}
\label{sec:radmc}
We have shown the impact of the dust grain sizes on the strength of the VSI and the morphology of the dust layer in the previous section.
Now, we want to determine the visual appearance of the simulated disks in synthetic \rev{millimeter-wave} observations. Our goal is a qualitative comparison of our results with ALMA observations of edge-on or almost edge-on protoplanetary disks. 
Specifically, the works of \cite{Villenave2020,Villenave2022,Villenave2023} have shown that many protoplanetary disks appear settled in $\lambda = \SI{1.25}{\milli \meter}$ images obtained with ALMA. Oph 163131 is the most prominent example with a very thin dust disk of height 
$H_\mathrm{d, \SI{100}{\AU}}\approx \SI{0.5}{\AU}$. \cite{Villenave2022} obtained this result by modeling the appearance of one of the disk's gaps.

For our approach, we create radiation intensity maps of edge-on disks ($i=\SI{90}{\degree}$) from \report{the dust distributions of the last snapshot of our hydrodynamic} simulations with \radmc{}. 
For comparison, we also simulate the intensities arising from steady-state dust distributions under the assumption of a fixed diffusivity. In this settling--mixing equilibrium, the vertical dust distribution can be written 
\begin{equation}
\label{eq:eps_vert}
\varepsilon = \varepsilon_\mathrm{mid} \exp\left[-\frac{\St_\mathrm{mid}}{\delta}\left(\exp\left(\frac{z^2}{2H_\mathrm{g}^2}\right) - 1\right)\right],
\end{equation}
\citep{Fromang2009}. Opacities are calculated for each of the four populations using the standard DSHARP particle properties with the \texttt{dsharp\_opac} python package \citep{Birnstiel2018}.
We consider a photon package to be fully extinct after being scattered over a length of five optical depths. Our models are axisymmetric and we treat the anisotropic scattering angle for \rev{60 angular sample points}. 
Before running the ray tracing algorithm, we use the \texttt{mctherm} task to calculate the dust temperatures from a thermal Monte Carlo simulation. For this, we use $10^7$ photon packages.

To mimic the effect of a finite beam size in ALMA observations, we convolve our images with a circular Gaussian beam, which for DSHARP observations had a typical FWHM of \SI{35}{mas}. We place our disk at a distance of \SI{100}{pc} to the observer. 

\begin{figure*}[ht]
    \centering
    \includegraphics{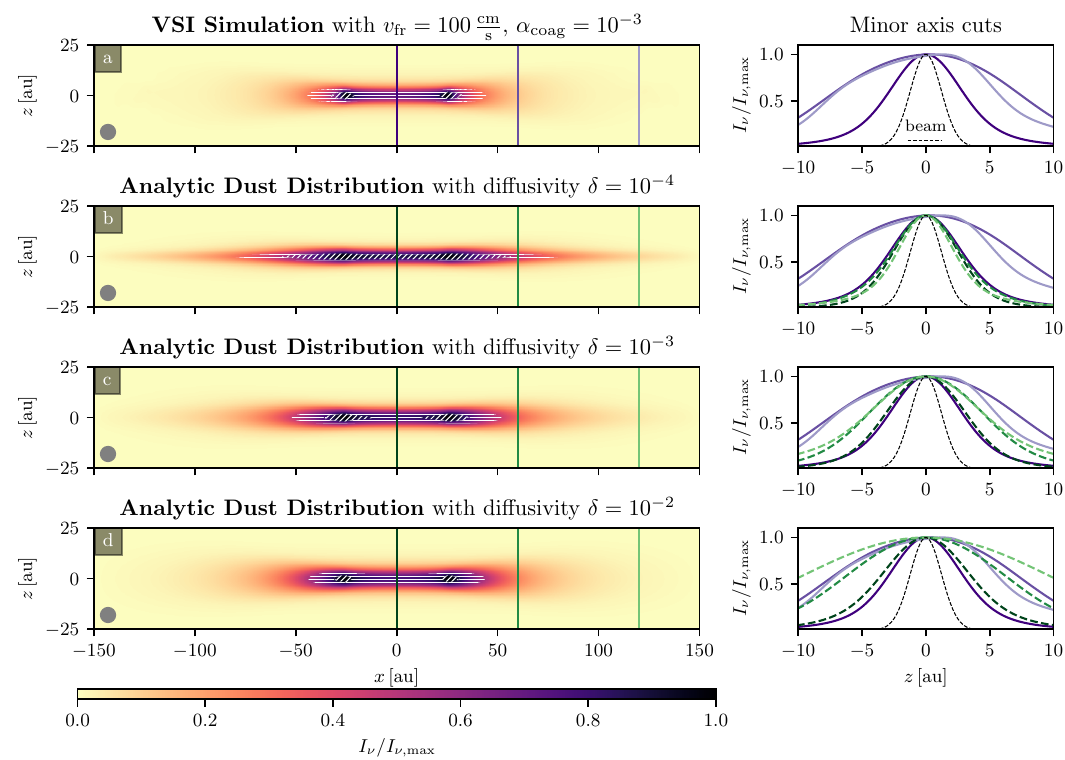}
    \caption{Upper row a: \radmc{} intensity maps of our VSI simulation with $v_\mathrm{fr}=\SI{100}{\centi \meter \per \second}$ and $\alpha=10^{-3}$, seen edge-on. 
    Rows b, c, and d show intensity maps calculated from analytic dust distribution that assume different diffusivities $\delta$. The grain sizes are identical in all simulations.
    We convolve the images with a typical ALMA beam with FWHM of \SI{35}{mas} for a distance of \SI{100}{pc} shown as a grey circle.   
    Hatched areas mark regions that have optical depth $\tau\geq 1$. Horizontal hatches correspond to areas for which the $\tau=1$ surface lies on the far side of the disk. 
    Diagonally hatched regions mark $\tau=1$ surfaces that lie on the observer's side of the disk.
    The panels on the right-hand side show minor axis cuts through the images along the vertical lines in the intensity maps. \report{Purple lines in all plots are the minor axis cuts from the VSI simulation (panel a).}}
    \label{fig:radmc100}
\end{figure*}

\begin{figure*}[ht]
    \centering
    \includegraphics{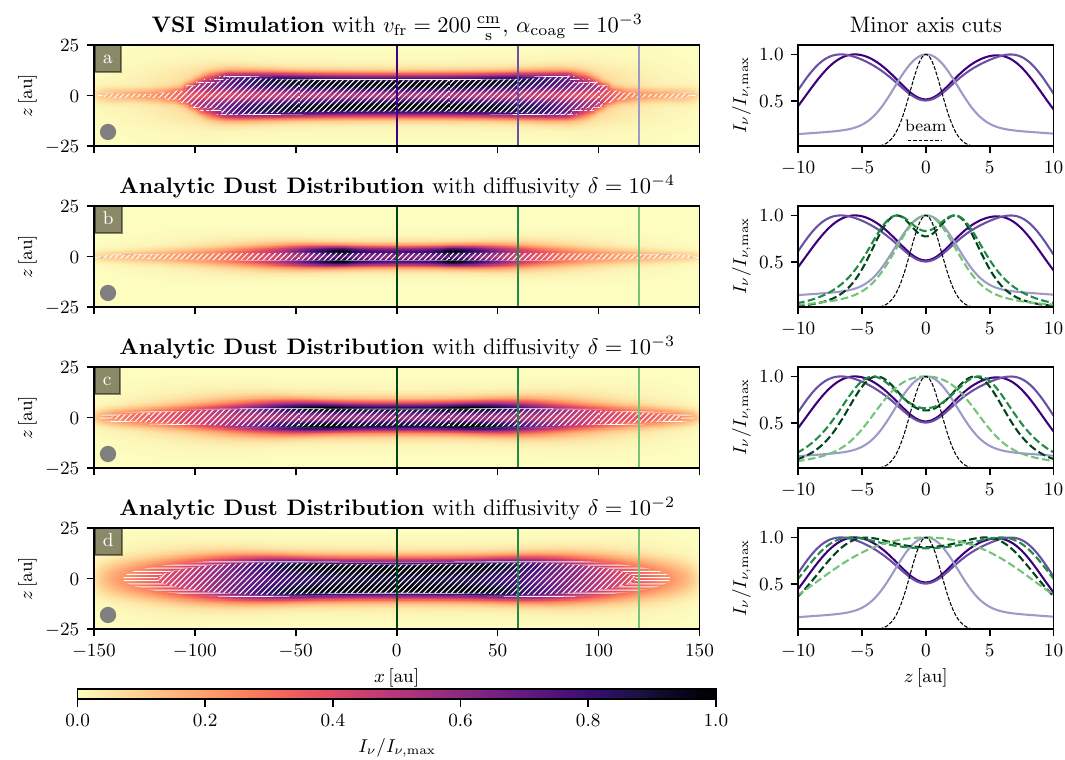}
    \caption{Upper row a: \radmc{} intensity maps of our VSI simulation with $v_\mathrm{fr}=\SI{200}{\centi \meter \per \second}$ and $\alpha=10^{-3}$, seen edge-on. 
    Rows b, c, and d show intensity maps calculated from analytic dust distribution that assume different diffusivities $\delta$. The grain sizes are identical in all simulations.
    We convolve the images with a typical ALMA beam with FWHM of \SI{35}{mas} for a distance of \SI{100}{pc} shown as a grey circle.   
    Hatched areas mark regions that have optical depth $\tau\geq 1$. Horizontal hatches correspond to areas for which the $\tau=1$ surface lies on the far side of the disk. 
    Diagonally hatched regions mark $\tau=1$ surfaces that lie on the observer's side of the disk.
    The panels on the right-hand side show minor axis cuts through the images along the vertical lines in the intensity maps. \report{Purple lines in all plots are the minor axis cuts from the VSI simulation (panel a).}}
    \label{fig:radmc200}
\end{figure*}

\begin{figure*}[ht]
    \centering
    \includegraphics{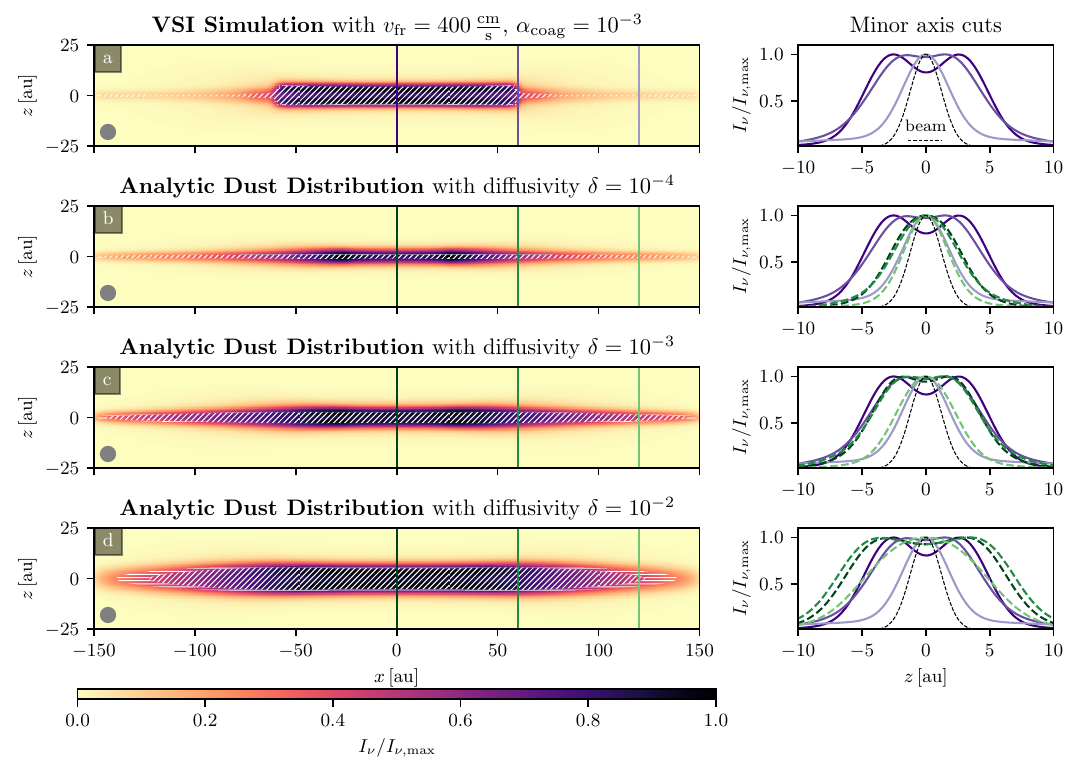}
    \caption{Upper row a: \radmc{} intensity maps of our VSI simulation with $v_\mathrm{fr}=\SI{400}{\centi \meter \per \second}$ and $\alpha=10^{-3}$, seen edge-on. 
    Rows b, c, and d show intensity maps calculated from analytic dust distribution that assume different diffusivities $\delta$. The grain sizes are identical in all simulations.
    We convolve the images with a typical ALMA beam with FWHM of \SI{35}{mas} for a distance of \SI{100}{pc} shown as a grey circle.   
    Hatched areas mark regions that have optical depth $\tau\geq 1$. Horizontal hatches correspond to areas for which the $\tau=1$ surface lies on the far side of the disk. 
    Diagonally hatched regions mark $\tau=1$ surfaces that lie on the observer's side of the disk.
    The panels on the right-hand side show minor axis cuts through the images along the vertical lines in the intensity maps. \report{Purple lines in all plots are the minor axis cuts from the VSI simulation (panel a).}}
    \label{fig:radmc400}
\end{figure*}

\begin{figure}
    \centering
    \includegraphics{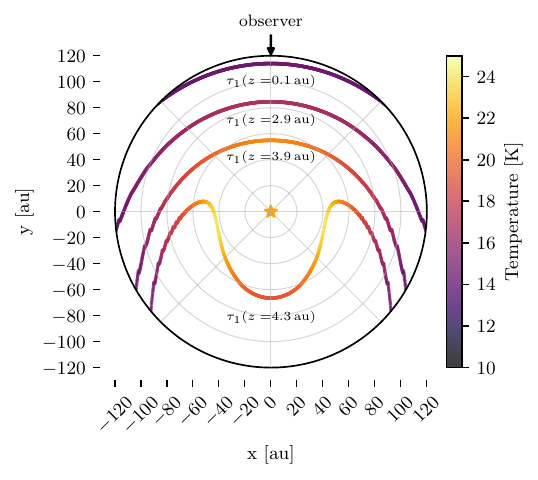}
    \caption{\rev{Origin of the double-peaked intensity profiles in \autoref{fig:radmc200} and \autoref{fig:radmc400}. The $\tau=1$ surfaces for the layers above the midplane lie closer to the central star due to the lower densities. The respectively higher temperatures lead to higher intensities above the midplane. Here shown are the $\tau=1$ surfaces for row c in \autoref{fig:radmc200}}.}
    \label{fig:polar}
\end{figure}

We show the resulting images for the VSI simulation with $v_\mathrm{fr}=\SI{100}{\centi \meter \per \second}$ in \autoref{fig:radmc100}, $v_\mathrm{fr}=\SI{200}{\centi \meter \per \second}$ in \autoref{fig:radmc200}, and for $v_\mathrm{fr}=\SI{400}{\centi \meter \per \second}$ in \autoref{fig:radmc400}. The right-hand side of each figure depicts three minor axis cuts through the intensity map at the locations of the vertical lines in the images.
The images within each figure are created from disk models with identical particle sizes.
As a result of optical depth effects, we find that the models with $v_\mathrm{fr}=\SI{200}{\centi \meter \per \second}$ (\autoref{fig:radmc200}) $v_\mathrm{fr}=\SI{400}{\centi \meter \per \second}$ (\autoref{fig:radmc400}), have a double-peaked intensity profile in the optically thick regions, marked by the hatched areas in each image.
Above the midplane, these models have optical surfaces closer to the central star. Therefore, we observe the hotter inner regions above the midplane and the cooler outer regions in the disk midplane, \rev{as illustrated in \autoref{fig:polar}}. 
\rev{Double-peaked profiles have already been observed in synthetic images of a VSI active disk in \cite{Blanco2021}. Their work is based on the simulation presented in \cite{Flock2020} and also treats radiative transfer through radiative diffusion in combination with ray-tracing from the central star for up to \SI{10}{\micro \meter} dust particles.}

The disk model with the smaller particles ($v_\mathrm{fr}=\SI{100}{\centi \meter \per \second}$), is subject to the strongest VSI and the strongest vertical mixing \rev{(row a of \autoref{fig:radmc100})}. Therefore, the disk midplane is not as strongly enriched and remains optically thin outside of $\sim \SI{45}{\AU}$. We are therefore not observing any double-peaked minor axis intensity profiles in these cases.
\rev{The minor cut intensity profiles in the inner disk match best with the analytic profile with $\delta=10^{-4}$ or $\delta=10^{-3}$ (rows b and c in \autoref{fig:radmc100}). In the outer disk, they show almost no settling, since the VSI is still active under the given conditions (more comparable with large diffusivities as in row c in \autoref{fig:radmc100})}.
Similar to the conclusions of \cite{Dullemond2022}, we confirm that such a disk structure is not consistent with observations of highly settled edge-on disks like Oph 163131.

In our disk model with $v_\mathrm{fr}=\SI{200}{\centi \meter \per \second}$, we find a vertically extended and optically thick inner disk with the typical two intensity peaks. However, we can already see the effect of the radially increasing cooling times in the regions beyond \SI{100}{\AU}. 
\rev{While the inner, VSI active regions appear the be more consistent with the analytic models of high diffusivity (row d in \autoref{fig:radmc200}), we can see that the outer regions are most consistent with a low diffusivity of $\sim 10^{-4}$ (row b in \autoref{fig:radmc200})}. This would still not be in agreement with observations of Oph 163131, which find the disk to be highly settled at $r\approx \SI{80}{\AU}$.

Ramping up the fragmentation threshold further, as in our model with $v_\mathrm{fr}=\SI{400}{\centi \meter \per \second}$, results in a \report{highly settled outer dust layer} outside of $\sim \SI{60}{\AU}$, as can be seen in \autoref{fig:radmc400}. The minor axis cuts illustrate the transition from an optically thick vertical structure in the inner regions to a \rev{mostly} optically thin profile in the outer regions, which occurs at the outer edge of the VSI active region. 
\rev{For the inner regions, we find a good agreement between the VSI simulation and the analytic model with $\delta=10^{-3}$ (row c in \autoref{fig:radmc400}).
Similar to \cite{Dullemond2022}, we find that the VSI can still lift up large particles in these inner regions.
In contrast, the outer regions are strongly settled, and more consistent with the analytic profile with $\delta=10^{-4}$ (row b in \autoref{fig:radmc400}). At this level of settling, it is unlikely that the outer regions of the VSI simulation could still be distinguished from a fully settled disk ($\delta=0$), due to the applied beam smearing.}
Note that in this simulation, we allow dust to flow into the simulation domain with a vertical distribution equal to the initial condition (which assumes $\delta=10^{-3}$). Any remaining vertical extent of the dust layer in the outer disk therefore likely exists as a result of the boundary condition. 

\section{Discussion}
\label{sec:discussion}
\subsection{Other Modes of Thermal Relaxation}
We assumed the dust to be the only source of cooling in the outer regions of protoplanetary disks. However, molecules like CO, H$_2$O, CO$_2$, etc, with electromagnetic dipole moments, might also contribute to the cooling of the gas through line emission when gas and dust are thermally decoupled \citep{Woitke2009, Malygin2014}. 
In this case, thermal energy must also be transferred from the bulk constituent of the disk, H$_2$, to the emitting species via collisions. Cooling the VSI modes could, thus, again become a matter of collision timescales at the very low densities of the outer disk. 
At low temperatures, emission lines may also become extremely inefficient at cooling the gas at the required rate. 
\hubert{Freeze-out of emitting molecules might also reduce the rate of thermal relaxation that can be achieved by emission line cooling. How much material can freeze out and thus be stopped from cooling the H$_2$, depends also on the availability of small grains. Cooling of the disk via gas emission lines is, thus, also dependent on the details of the dust population.}
Future studies should aim to incorporate some treatment of gas cooling via emission lines. Models for this exist \citep{Woitke2015}, but are very complex and currently not feasible for implementation in a hydrodynamic simulation.

Furthermore, we have omitted the optically thick regions of protoplanetary disks ($R<\SI{10}{\AU}$) in our simulations. Optically thick in this context does not refer to the bulk optical depth \rev{of the disk $(\tau\sim\Sigma \kappa)$}, as discussed in the previous section, but to the optical depth of individual VSI flow structures, which in the inner disk measure only a fraction of the disk scale height in the radial direction\rev{(denoted as $l$ in the following)}. We attempted to simulate these regions in \cite{Pfeil2021} by assuming a characteristic diffusion length scale. However, self-consistent modeling requires some treatment of radiative transfer, as in \cite{Stoll2016} or \cite{Flores-Rivera2020}.
Our findings nonetheless allow us to make predictions about the effect of dust coagulation on the cooling times in these regions, based on the results obtained here.
If radiative diffusion becomes the dominant channel for thermal relaxation, we can write the respective cooling time as
\begin{equation}
    t_\mathrm{LTE}^\mathrm{diff}= \frac{3}{16}\frac{C_V\rho_\mathrm{small}\rhog \kappa_\mathrm{R} l^2}{\sigma_\mathrm{SB}T^3},
\end{equation}
where $\kappa_\mathrm{R}$ is the Rosseland mean opacity, which is mostly determined \rev{by the small grains of density} $\rho_\mathrm{small}$ \citep{Lin2015, Dullemond2022}.
If coagulation is increasing the maximum particle size, the density of small particles will be reduced, therefore reducing the diffusion timescale. 
At the same time, the size distribution-averaged opacity will also be reduced. 
Therefore, dust coagulation would effectively reduce the diffusion time scale and thus be beneficial for the VSI in the inner disk regions. 

\subsection{Implication for the Vertical Shear Instability}
We have shown that the vertical shear instability is highly sensitive to the underlying dust size distribution, which determines the timescale of thermal relaxation.
\cite{Manger2021} \rev{and \cite{Klahr2023}} have shown that the VSI \rev{growth rate} almost instantaneously drops to almost zero once the critical cooling time threshold is reached. This is also what we observe as a sudden cutoff in the VSI activity at large disk radii.
Therefore, the VSI active zones in protoplanetary disks are not extending throughout the entire outer disk. 
Our simulations predict a VSI dead zone at large radii, which is caused by the reduced efficiency of cooling.

Our simulations omit a treatment of dust back-reaction onto the gas. \cite{Schafer2020} have shown, that if the dust can settle into a thin layer in the disk midplane before the VSI starts to grow, dust feedback can counteract the VSI. Since dust coagulation, settling, and the onset of the VSI, occur on comparable timescales, it is not trivial to predict the outcome of such a situation without a realistic disk simulation that treats all of the aforementioned effects simultaneously.
Our results show that if some dust settling and coagulation can occur before the onset of the VSI, the effect of the reduced cooling time would reduce the VSI activity and therefore probably enhance the dampening effect of the dust's dynamic back-reaction onto the gas.

\subsection{\rev{The Need for a Self-consistent Three-dimensional Model and the Limitations of Our Approach}}
Simulations that aim to study the VSI under realistic conditions cannot ignore the implications of an evolved dust population, as presented in our and previous studies \citep[see][]{Fukuhara2021, Fukuhara2023}. 
Measurements of the spectral index in protoplanetary disks \citep{Tazzari2016, Perez2012, Huang2018, Sierra2021} and polarization observations \citep{Ohashi2019} imply that dust coagulation is occurring and that grains in the outer disk can reach sizes of between \SIrange[range-units = single]{0.1}{1}{\milli\meter}, similar to the outcome of the \dpy{} models that our VSI simulations are based on.
Note, however, that our studies are no self-consistent representations of protoplanetary disks. 
The dust size distributions used to calculate the cooling time in our setups are static. In a real disk, they would evolve together with the VSI. Settling and stirring of the dust layer would impact the cooling times. It is unclear if this would lead to some sort of equilibrium situation in which the dust stirring by the VSI can maintain a thick enough dust layer to support the necessary cooling times. Continuous coagulation of grains would counteract the turbulent mixing further.

\cite{Fukuhara2023} presented an approach to study this equilibrium by using analytic, yet physically motivated, cooling time profiles. They iterated between VSI simulations and calculations of the resulting steady-state dust distribution from the measured turbulent diffusivity. In that way, they were able to reach a convergent state in which the VSI turbulence creates the necessary diffusivity to maintain the underlying cooling times. 
Their studies did, however, not consider the effect of the changing diffusivity on the grain size itself through coagulation and fragmentation. This poses an additional uncertainty in their and our studies.
\hubert{We can already see that the measured Mach numbers in our simulations do not always correspond to the $\alpha$ values used in the underlying coagulation models (see \autoref{fig:MachTimeVertical}). Note that $\mathcal{M}$ is only part of the generation of turbulent collision speeds \citep{Ormel2007}. The turbulent spectrum in correlation time space is also required to calculate the acceleration that can be imposed on various particle sizes. Collision speeds can only be obtained from the large scale rms\ velocity $U(L)$ and the associated length scale $L = \sqrt{\alpha} H$, for an ideal Kolmogorov turbulence cascade which causes isotropic turbulent diffusivities \citep{Youdin2007, Binkert2023}.}

\report{If any source of additional turbulence would be present that causes the turbulent diffusivities used in our coagulation models, this would also have an effect on the developing VSI. Even small viscosities of $\alpha=\SIrange{1e-4}{1e-3}{}$ are enough to hinder the evolution of the VSI \citep{Barker2015}. Future studies should try to apply a more realistic, self-consistent prescription of diffusivities in the coagulation model.}

\rev{In our cooling time calculations, we have also neglected the effects of radial drift. Drift-limited size distributions are characterized by smaller maximum particle sizes and are more top-heavy than fragmentation-limited distributions. This results in longer thermal accommodation timescales and would further inhibit the VSI turbulence.}

\rev{The effect of the drag force onto the gas was also not considered in our simulations. \cite{Schafer2020} and \cite{Schafer2022} have shown that back-reaction can indeed inhibit the VSI turbulence close to the disk midplane if the dust has time to sediment before the VSI is saturated. Future studies should therefore aim to incorporate more realistic dust dynamics.}

\hubert{In our two-dimensional simulations with dust, we have observed flat top or double-peaked dust-to-gas ratio distributions. This reflects the periodic and non-isotropic nature of the VSI-driven turbulence\report{, which is not accounted for in the coagulation simulations}. However, as our simulations are two-dimensional, the prominence of these features might be artificially enhanced, as the $\varphi-$dimension is missing as a degree of freedom. Three-dimensional simulations \citep{Manger2018,Flock2020,Manger2021,Pfeil2021} are needed for the study of the non-linear saturation and fully developed turbulent state of VSI-driven turbulence, before deriving the turbulence properties as diffusivity, correlation times, and energy spectra.}

The main conclusions of our study and \cite{Fukuhara2021}, however, remain unchanged by all these considerations.
Dust coagulation and dynamics are essential components in studies of cooling-time-sensitive instabilities like the VSI.

This highlights the need for a more-self consistent numerical approach. Cooling times have to be constantly recalculated throughout a simulation from the present dust size distributions in order to study such systems. 
In the inner, optically thick parts of the disk, radiative transfer models have to be employed to study the effect of coagulation on diffusive radiative cooling.

\section{Summary and Conclusions}
In this work, we studied the effect of evolved dust size distributions on the VSI activity in protoplanetary disks. We conducted hydrodynamic simulations based on five different dust coagulation models for different fragmentation velocities and assumed turbulence strengths, which resulted in maximum particle sizes between $\sim\SI{10}{\micro \meter}$ and $\sim\SI{0.1}{\centi \meter}$.
Based on these dust size distributions, we calculated the cooling times for our subsequent hydrodynamic simulations.
Our results show a strong dampening effect of dust coagulation on the VSI, as predicted by previous studies \citep{Lin2015, Fukuhara2021, Pfeil2021, Dullemond2022, Fukuhara2023}. The reason for this is the collisional decoupling between dust particles and gas molecules that is enhanced if dust coagulation is increasing the maximum particle size. Reduced collision rates inhibit the thermal accommodation of dust and gas and therefore reduce the cooling rate of the gas.

The effect can be strong enough to \report{hinder the development of the VSI, leading to a highly settled dust layer} even for moderate fragmentation velocities of $v_\mathrm{fr}\gtrsim\SI{200}{\centi\meter\per\second}$.
At the same time, the inner regions---in which the gas and dust components remain well coupled---can maintain some level of VSI turbulence.
This finding is consistent with recent observations of highly settled dust layers in protoplanetary disks \citep{Villenave2020, Villenave2022}. 
Our simulations also show that even a low level of VSI can still significantly alter the vertical distribution of dust, which we can observe in the inner disk regions of our simulations with the largest particles. Synthetic millimeter-images of these VSI active regions are mostly consistent with analytic models that assume large diffusivities of $\delta\sim\SIrange{e-3}{e-2}{}$. At the same time, outer disk regions can appear completely settled in our simulations.
We thus report the existence of a VSI dead zone in the outer regions of protoplanetary disks. 
\rev{The existence of the VSI dead zone in the outer regions of protoplanetary disks reconciles recent millimeter-wave observations with models of hydrodynamic turbulence.}

\hubert{Future studies of VSI-active disks should aim to incorporate a more self-consistent treatment of dust coagulation and dynamics. Additionally, cooling via gas emission lines has to be considered to gain a better understanding of the impact of thermal relaxation on the VSI in protoplanetary disks. For this, thermochemical modeling is required to track the amounts and the evolution of relevant species, which in fact also depends on the dust coagulation process.
Modeling the optically thick parts of protoplanetary disks and the impact of stellar irradiation furthermore requires radiative transfer modeling.}

After applying our methodology to smooth disks in this article, we will extend our studies to disks with substructure in Part II. Specifically, Oph 163131 \citep{Villenave2020, Wolff2021, Villenave2022} and HD 163296 \citep{Dullemond2018, Rosotti2020, Doi2021} have been extensively surveyed with a focus on the dust diffusivities and provide good conditions for comparison with simulations.  


\section*{Acknowledgments}
T.P., H.K., and T.B. acknowledge the support of the German Science Foundation (DFG) priority program SPP 1992 “Exploring the Diversity of Extrasolar Planets” under grant Nos.\ BI 1816/7-2 and KL 1469/16-1/2. 
T.B. acknowledges funding from the European Research Council (ERC) under the European Union’s Horizon 2020 research and innovation programme under grant agreement No 714769 and funding by the Deutsche Forschungsgemeinschaft (DFG, German Research Foundation) under grants 361140270, 325594231, and Germany’s Excellence Strategy - EXC-2094 - 390783311.
Computations were conducted on the computational facilities of the University of Munich (LMU).

We thank the anonymous referee for their constructive comments, which helped us to improve the quality of this article.

\section*{Software}
\begin{itemize}
    \setlength{\itemsep}{0pt}
    \setlength{\parskip}{0pt}
    \item \href{http://plutocode.ph.unito.it/}{\pluto{}\texttt{-4.4}} \citep{Mignone2007}
    \item \href{https://www.ita.uni-heidelberg.de/~dullemond/software/radmc-3d/}{\radmc{}} \citep{Dullemond2012} 
    \item \href{https://www.python.org/}{\texttt{Python}} with the packages:
    \begin{itemize}
    \setlength{\itemsep}{0pt}
    \setlength{\parskip}{0pt}
        \item \href{https://numpy.org/}{\texttt{NumPy}} \citep{Harris2020}
        \item \href{https://scipy.org/}{\texttt{SciPy}} \citep{Virtanen2020}
        \item \href{https://matplotlib.org/}{\texttt{matplotlib}} \citep{Hunter2007}
        \item \href{https://github.com/stammler/dustpy}{\dpy{}} \citep{Stammler2022} 
        \item \href{https://github.com/stammler/dustpylib}{\texttt{DustPyLib}}
        \item \href{https://github.com/birnstiel/dsharp_opac}{\texttt{dsharp\_opac}} \citep{Birnstiel2018a}
        \item  \href{https://www.ita.uni-heidelberg.de/~dullemond/software/radmc-3d/manual_rmcpy/index.html}{\texttt{RADMC-3DPy}}
    \end{itemize}
\end{itemize}

\clearpage
\bibliography{Literature}

\appendix
\section{Cooling Time Derivations by Barranco et al. (2018)}
\label{sec:Appendix_Derivation}
To calculate our cooling times, we follow the derivations by \cite{Barranco2018}. 
We assume that the emission of dust grains determines the thermal relaxation of the gas in a two-stage process. Thermal energy is transferred between gas and dust molecules through collisions. The dust, which at low temperatures typically has much higher emissivity, can radiate an excess of thermal energy and thus effectively cool the gas.

\paragraph{Dust Emission Timescale}
This is the timescale on which the dust grains of temperature $T_\mathrm{d}$ reach thermal equilibrium with their surroundings (of temperature $T_\mathrm{eq}$, set by stellar irradiation in our model) via emission or absorption of electromagnetic radiation. We assume the dust grains to radiate as black bodies, meaning the cooling rate per unit volume of material can be derived by integrating the cooling rate per gram of dust of size $a$, over the size distribution
\begin{align}
    \Lambda_\mathrm{rad}^{\mathrm{d}} &=\int_{\amin}^{\amax} \, \rho(a) \Lambda_\mathrm{rad}(a) \mathrm{d}a \\
    &=  4 \sigma_\mathrm{SB} (T_\mathrm{d}^4-T_\mathrm{eq}^4)  \int_{\amin}^{\amax} \rho(a)\kappa_\mathrm{P}(a,T_\mathrm{eq}) \, \mathrm{d}a
\end{align}
The respective cooling timescale follows from 
\begin{align}
    t_\mathrm{d}^\mathrm{rad} & = \frac{\rhod C_\mathrm{d}}{\Lambda_\mathrm{rad}^{\mathrm{d}}}|T_\mathrm{d}-T_\mathrm{eq}| \\
    &=\frac{\rhod C_\mathrm{d}}{4 \sigma_\mathrm{SB} \displaystyle} \left(\frac{T_\mathrm{d}-T_\mathrm{eq}}{T_\mathrm{d}^4-T_\mathrm{eq}^4}\right) \left(\int_{\amin}^{\amax} \rhod(a)\planck(a,T_\mathrm{eq}) \, \mathrm{d}a\right)^{-1} 
\end{align}
Expanding this expression in a Taylor series for small $\delta T = T_\mathrm{d}-T_\mathrm{eq}$, results in 
\begin{equation}
    t_\mathrm{d}^\mathrm{rad} \approx \frac{\rhod C_\mathrm{d} }{\displaystyle 16 \,\sigma_\mathrm{SB}\, T_\mathrm{eq}^3} \left(\int_{\amin}^{\amax} \rhod(a)\planck(a,T_\mathrm{eq}) \, \mathrm{d}a\right)^{-1},
\end{equation}

\paragraph{Thermal Accommodation of the Gas via Collisions with the Dust}
The emission or absorption of radiation by the dust grains can relax temperature perturbations in the gas only if dust and gas are thermally coupled via collisions. We calculate this collisional accommodation timescale by integrating the collision rate per unit volume over the entire size distribution, assuming a thermal accommodation coefficient $\mathcal{A}=0.5$, as in \cite{Barranco2018}. 
In this case, the thermal energy transferred between gas and dust can be written as
\begin{align}
    \Lambda_\mathrm{coll} &= \int_{\amin}^{\amax} n_\mathrm{d}(a)  \sigma_\mathrm{coll}  \bar{v}_\mathrm{g} n_\mathrm{g} 
 \, 2\mathcal{A}k_\mathrm{B}(T_\mathrm{g}-T_\mathrm{d})\, \mathrm{d}a  \\
 &=   \pi  \bar{v}_\mathrm{g} n_\mathrm{g}
 \, k_\mathrm{B}(T_\mathrm{g}-T_\mathrm{d})\, \int_{\amin}^{\amax} n(a) a^2 \mathrm{d}a,
\end{align}
where $n_\mathrm{g}$ is the number density of gas molecules \report{and $T_\mathrm{g}$ is the gas temperature} \citep{Probstein1969, Burke1983}.
Now we multiply by $\nicefrac{\rho_\mathrm{d}}{\rho_\mathrm{d}}$ where the total dust density can be written $\rho_\mathrm{d}=\int n(a) m(a) \, \mathrm{d}a$, with the particle mass $m(a)=\nicefrac{4}{3} \, \pi a^3 \rho_\mathrm{m}$. Thus, we obtain
\begin{align}
     \Lambda_\mathrm{coll} &=   \bar{v}_\mathrm{g} n_\mathrm{g} 
 \, k_\mathrm{B}(T_\mathrm{g}-T_\mathrm{d})\frac{3\rho_\mathrm{d}}{4 \rho_s}\, 
 \frac{\int_{\amin}^{\amax} n(a) a^2 \mathrm{d}a}{\int_{\amin}^{\amax} n(a) a^3 \mathrm{d}a}
\end{align}
Here, we can insert the definition of the Sauter mean radius $a_\mathrm{s}=\nicefrac{\langle a^3\rangle}{\langle a^2\rangle}$, where $n_\mathrm{d}\langle a^2\rangle \coloneqq \int_{\amin}^{\amax} n(a) a^2 \mathrm{d}a$ is the second moment of the size distribution and $n_\mathrm{d}\langle a^3\rangle \coloneqq \int_{\amin}^{\amax} n(a) a^3 \mathrm{d}a$ is the third moment of the size distribution.
With this, we arrive at Equation 12 from \cite{Barranco2018}
\begin{equation}
\label{eq:Barranco12}
    \Lambda_\mathrm{coll} = \left(\frac{3}{4\rho_\mathrm{m}}\right)\left(\frac{1}{a_\mathrm{s}}\right)\left(\frac{\rho_\mathrm{d}}{\rho_\mathrm{g}}\right)\left(\frac{\rho^2_\mathrm{g}\bar{v}_\mathrm{g}}{\bar{m}_\mathrm{g}}\right) k_\mathrm{B}(T_\mathrm{g}-T_\mathrm{d}).
\end{equation}
The collisional cooling time of the gas via collisions with the dust follows from 
\begin{equation}
    t_\mathrm{g}^\mathrm{coll} = \frac{\rhog C_P}{\Lambda_\mathrm{coll}}|T_\mathrm{g}-T_\mathrm{d}|.
\end{equation}
We insert \autoref{eq:Barranco12} and multiply by $\nicefrac{\pi  a_\mathrm{S}^3}{\pi a_\mathrm{S}^3}$. For particles of size $a_\mathrm{S}$, we define a hypothetical number density of $n_\mathrm{S}=\nicefrac{\rhod}{m_\mathrm{S}}$ and a collisional cross-section $\sigma_\mathrm{S}=\pi a_\mathrm{S}^2$. 
We apply Meyer's relation $\nicefrac{k_\mathrm{B}}{\bar{m}}=C_P-C_V$ and the definition of the heat capacity ratio  $\gamma=\nicefrac{C_P}{C_V}$ and arrive at \autoref{eq:tgcoll}
\begin{equation*}
        t_\mathrm{g}^\mathrm{coll} = \frac{\gamma}{\gamma-1}\frac{1}{n_\mathrm{S}\sigma_\mathrm{S}\Bar{v}_\mathrm{g}}.
\end{equation*}
Likewise, the dust grains adjust their temperature on a timescale
\begin{align}
        t_\mathrm{d}^\mathrm{coll} &= \frac{\rhod C_\mathrm{d}}{\Lambda_\mathrm{coll}}(T_\mathrm{g}-T_\mathrm{d}). \\
        &= \left(\frac{\rhod}{\rhog}\right)\left(\frac{C_\mathrm{d}}{C_P}\right)t_\mathrm{g}^\mathrm{coll},
\end{align}
where $C_\mathrm{d}$ is the specific heat capacity of the dust grains, which we take to be \SI{800}{\joule \per \kelvin \per \kilo\gram} \citep[see][]{Barranco2018}.
We note that for a truncated power law size distribution $n(a)\propto a^p$, the Sauter mean radius can be written
\report{
\begin{align}
a_\mathrm{S} =
\begin{cases}
    \frac{\amax-\amin}{\log(\amax)-\log(\amin)} &\text{for $p=-3$} \\
    \frac{\amax\amin}{\amax-\amin} \log\left(\frac{\amax}{\amin}\right)&\text{for $p=-4$} \\
    \left(\frac{p+3}{p+4}\right) \frac{\amax^{p+4}-\amin^{p+4}}{\amax^{p+3}-\amin^{p+3}} &\text{for $p\neq -4,-3$}
\end{cases}
\end{align}
}
which, for a typical size distribution with $p=-3.5$, leads to to $a_\mathrm{S}=\sqrt{\amax\amin}$.
The collisional cooling time of the gas (\autoref{eq:tgcoll}), therefore scales as $t_\mathrm{g}^\mathrm{coll}\propto \sqrt{\amax} $. If the size distribution is fragmentation limited, which is to expected for most parts of the disk, this translates to
\begin{equation}
    t_\mathrm{g}^\mathrm{coll} \propto \frac{v_\mathrm{fr}}{\sqrt{\alpha}}.
\end{equation}

\paragraph{Total Thermal Relaxation Time of the Gas}
In order to derive the total thermal relaxation time of the gas, \cite{Barranco2018} write the equations of thermal relaxation as
\begin{align}
    \frac{\mathrm{d} (\delta T_\mathrm{g})}{\mathrm{d}t} &= -\frac{\delta T_\mathrm{g}-\delta T_\mathrm{d}}{t_\mathrm{g}^\mathrm{coll}} \\
    \frac{\mathrm{d} (\delta T_\mathrm{d})}{\mathrm{d}t} &= \frac{\delta T_\mathrm{g}-\delta T_\mathrm{d}}{t_\mathrm{d}^\mathrm{coll}} - \frac{\delta T_\mathrm{d}}{t_\mathrm{d}^\mathrm{rad}}.
\end{align}
For this coupled system, one can look for solution of the form $\delta T_\mathrm{g}= \hat{T}_\mathrm{g}\exp(-t/t_\mathrm{thin}^\mathrm{NLTE})$, $\delta T_\mathrm{d}=\hat{T}_\mathrm{d}\exp(-t/t_\mathrm{thin}^\mathrm{NLTE})$.
The resulting equation system can be solved for $t_\mathrm{thin}^\mathrm{NLTE}$, which results in \autoref{eq:tNLTE}.

\section{Cooling Time Maps}
\label{sec:Appendix_Fits}

To make our cooling time maps, derived from \dpy{} models, usable for our \pluto{} simulation, we fitted the data as a function of gas density and temperature. In this way, we can easily calculate the cooling time for each grid cell in \pluto{} from the local disk structure.

We found that for a given temperature, the cooling time can be fitted by a power law in density,
\begin{equation}
t_\mathrm{NLTE, Fit} = t_\mathrm{10^{-15}}(T)\left(\frac{\rho_\mathrm{g}}{\SI{e-15}{\gram \per\cubic \centi\meter}} \right)^{p(T)}
\end{equation}
The power law's parameters themselves can be fitted as broken broken power laws of temperature:
\begin{align}
    t_\mathrm{10^{-15}}(T) &= \left(\frac{T}{T_{t, \mathrm{c}}}\right)^{q_t+s_t} \frac{T_{t, \mathrm{c}}^{s_t}}{(T_{t, \mathrm{c}}^{s_t}+T^{s_t})} + p_t \\
    p(T) &= \left(\frac{T}{T_{p, \mathrm{c}}}\right)^{q_p+s_p} \frac{T_{p, \mathrm{c}}^{s_p}}{(T_{p, \mathrm{c}}^{s_p}+T^{s_p})} + p_p
\end{align}

\begin{figure*}
    \centering
    \includegraphics[width=1.0\textwidth]{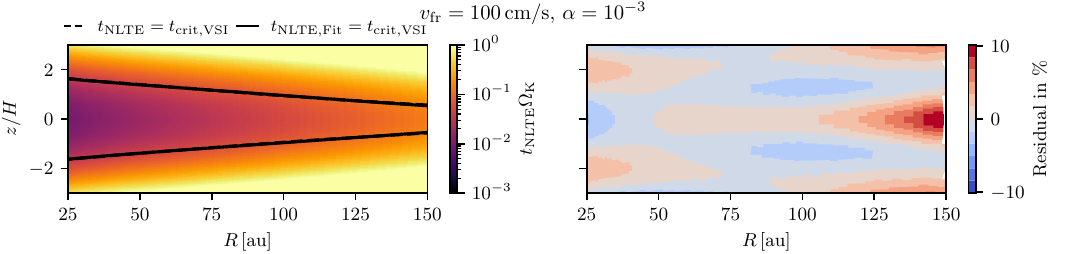}
    \includegraphics[width=1.0\textwidth]{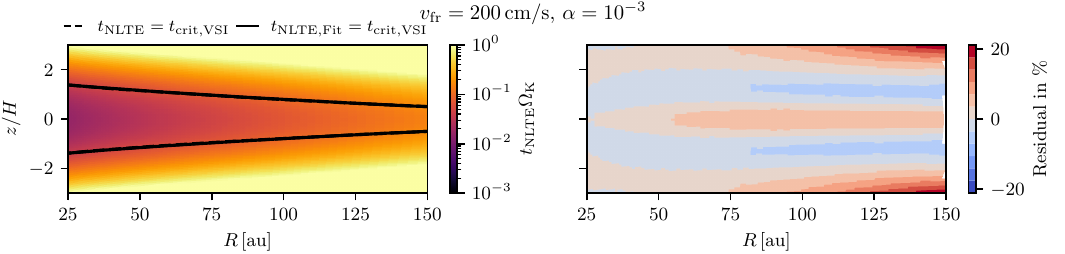}
    \includegraphics[width=1.0\textwidth]{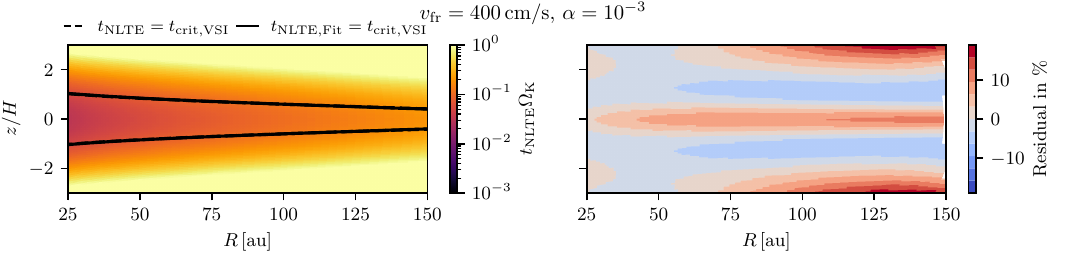}
    \includegraphics[width=1.0\textwidth]{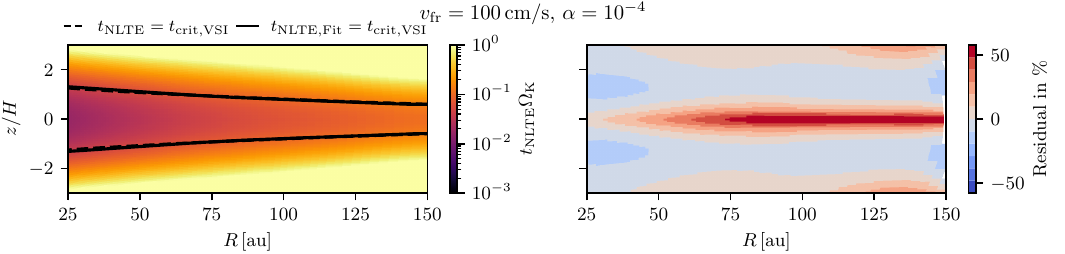}
    \includegraphics[width=1.0\textwidth]{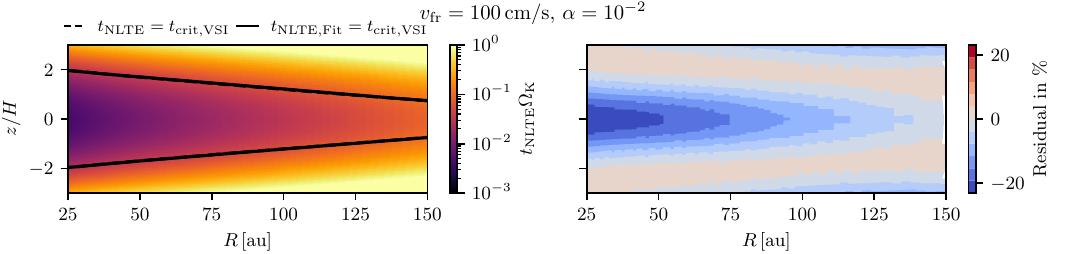}
    \caption{Cooling time maps in our 5 VSI \pluto{} simulations on the left-hand side, and the residuals between the fit used in our simulations, and the actual cooling time distributions derived from \dpy{} simulations.}
    \label{fig:CoolingFits}
\end{figure*}

This means that our two-dimensional cooling time distribution can be described as a function of $\rho_\mathrm{g}$ and $T$ with a total of eight parameters.
For the actual fitting procedure, we use the \texttt{scipy} routine \texttt{curve\_fit}.
Residuals between the actual cooling time maps and our fitting functions are displayed in \autoref{fig:CoolingFits}. With the exception of our model with $v_\mathrm{fr}=\SI{100}{\centi \meter \per \second}$ and $\alpha=10^{-4}$, all fits have maximum deviations from the data of $<\SI{30}{\percent}$ in the entirety of the simulation domain. The respective model with higher deviations corresponds to a highly settled particle layer with $\alpha=10^{-4}$, making it distinct from the other models with $\alpha=10^{-3}$. Thus, some deviation had to be expected for this case.
Note that we are generally interested in trends of the VSI activity with the given coagulation parameters, which are well captured by our fits. 
Uncertainties of \SIrange{10}{60}{\percent} in the cooling times thus do not influence the conclusions of our work.

\section[Dust Advection and Diffusion in PLUTO]{Dust Advection and Diffusion in \pluto{}}
\label{sec:Appendix_Tests}

As a consequence of the sub-Keplerian azimuthal gas velocity, particles in aerodynamic force equilibrium with the gas also have sub-Keplerian terminal velocities. Radial pressure forces, which contribute the to gas' radial force balance, do not significantly act on the dust grains. Therefore, the grains embedded in the gas cannot stay on circular orbital trajectories and spiral inward at a given terminal drift speed.
\cite{Nakagawa1986} derived this radial drift velocity as
\reporttwo{
\begin{equation}
    v_\mathrm{d-g, r}= v_{\mathrm{d},r} - v_{\mathrm{g},r} = \frac{\St (1+\varepsilon)}{\St^2 + (1+\varepsilon)^2}\frac{1}{\rhog\Omega_\mathrm{K}}\frac{\partial P}{\partial R}  \approx \frac{\St}{\St^2+1}\frac{1}{\rho_\mathrm{g}\Omega_\mathrm{K}}\frac{\partial P}{\partial R},
    \label{eq:vdrift}
\end{equation}
where in our simulations $P$ is the gas pressure, and therefore subject to fluctuations arising from the VSI. We use the zeroth order approximation for small dust-to-gas ratios on the right-hand side. This approximation is robust in the VSI-active regions, where the dust-to-gas ratios are generally smaller than 0.05 in our simulations.} 
A similar derivation can be made for the vertical velocity component. Pressure forces keep the gas on elevated trajectories around the central star (acting against the vertical component of the gravitational force). Again, these forces \rev{have a negligible effect on the grains}. An expression, equal to \autoref{eq:vdrift}, can be found for the settling velocity of the grains
\begin{equation}
\label{eq:vset}
    v_\mathrm{d-g, z} = v_{\mathrm{d},z} - v_{\mathrm{g},z} = \frac{\St}{\St^2+1}\frac{1}{\rho_\mathrm{g}\Omega_\mathrm{K}}\frac{\partial P}{\partial z}.
\end{equation}

The \pluto{} code already allows for the treatment of passive tracer fluids, which are simply advected with the gas following
\begin{equation}
    \der{(\rhog \varepsilon)}{t} + \vec{\nabla} \cdot (\varepsilon \,  \rhog \, \vg) = 0.
\end{equation}
In our case, the advected quantity $\varepsilon$ represents a local dust-to-gas ratio $\varepsilon =\rhod/\rhog$. In the short friction time, terminal velocity approximation, the respective dust flow is modified to simulate a dust fluid that is aerodynamically coupled to the gas, i.e., undergoes radial and azimuthal drift, and vertical sedimentation, with terminal velocities given by \autoref{eq:vdrift} and \autoref{eq:vset}. 
to the tracer flux, which is valid for small dust-to-gas ratios \citep{Youdin2005}.
The tracer flux is then calculated with the upstream dust density based on the above velocity at the respective cell interface
\begin{equation}
\label{eq:dr_flux}
\vec{F}_\mathrm{drift, interface} =\rho_\mathrm{d, upstream} \vec{v}_\mathrm{d-g, interface}.
\end{equation}

\subsection{Test Case for Dust Drift}
\cite{Youdin2002}, presented an analytic description for the time evolution of a dust fluid with a fixed grain size in a protoplanetary disk due to radial drift. Their prescription was further developed in \cite{Birnstiel2014}, and their general solution to the advection equation is given by 
\begin{equation}
    \Sigma(r,t)=\Sigma(r_0,t)\frac{v(r_0)r_0}{v(r)r},
\end{equation}
with the velocity $v(r)$ and the original location of the characteristic $r_0$, defined via
\begin{equation}
    \frac{t}{r_\mathrm{c}} = \int_{r_0}^r \frac{1}{u(r')}\, \mathrm{d}r'
\end{equation}

\begin{figure}
    \centering
    \includegraphics{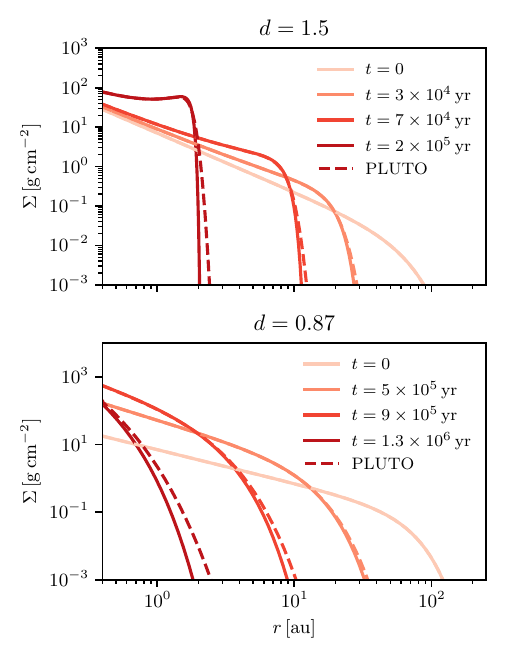}
    \caption{Radial dust advection tests based on \cite{Youdin2002}.}
    \label{fig:YoudinShu2002}
\end{figure}

Based on the assumption that the transport velocity scales as a radial power law $v_\mathrm{dr}\propto R^d$, one can solve the above integral and, thus, write a complete solution for the transport equation as
\begin{equation}
    \Sigma(r,t) = \Sigma_0 r^{-d-1} r_0^{d+1-\beta_\Sigma},
\end{equation}
with the time dependence entering through the initial radius of the characteristic 
\begin{equation}
    r_0 = r\left[1-(d-1)\frac{v_\mathrm{dr}t}{r}\right]^{-\frac{1}{d-1}}.
\end{equation}
We are using this analytic solution of the advection equation to verify our transport scheme for the \pluto{} code with a passive dust fluid.
We present two cases for different drift parameters $d$ in \autoref{fig:YoudinShu2002} for a particle size of \SI{0.1}{\centi \meter}. As can be seen, our transport scheme provides good agreement with the analytic transport solutions even for long integration times. However, we can also see that our simple donor-cell flux prescription is associated with a certain numerical diffusivity which is especially evident over very long integration times, as in the lower panel of \autoref{fig:YoudinShu2002}.

\subsection{Tests for Dust Diffusion (not used in VSI simulations)}
For completeness, we also introduce a diffusion velocity to our dust advection model, determined by the gradient in dust-togas ratio $\varepsilon$ and the diffusion coefficient $D$ 
\begin{equation}
    \vec{v}_\mathrm{diff}=-D\Vec{\nabla}\ln{(\varepsilon)},
\end{equation}
which we add to the advection velocity before the calculation of the complete upstream transport flux. We do not employ any additional diffusivity in our VSI simulations, meaning $\vec{v}_\mathrm{diff}=0$ in all presented simulation.
To test this very simple approach to dust diffusion in the \pluto{} code, we run two simulations with dust species of fixed Stokes numbers that reach a steady state that can be compared to steady-state solutions of the diffusion equation.
Our first test case considers dust trapping in a radial pressure bump and a fixed diffusivity parameter $\delta$.
\cite{Dullemond2018}, considered this scenario and derived an approximate analytic solution for a Gaussian pressure bump of amplitude $P_0$ and width $w$,
\begin{equation}
    P(r)=P_0\exp\left(-\frac{(r-r_0)^2}{2w^2}\right).
\end{equation}
This results in a steady state dust distribution, given by 
\begin{equation}
    \Sigma_\mathrm{d} = \Sigma_\mathrm{d,0}\exp\left(-\frac{(r-r_0)^2}{2w_\mathrm{d}^2}\right),
\end{equation}
where the width of the dust density distribution is related to the pressure bump's width via
\begin{equation}
    w_\mathrm{d}=w\sqrt{\frac{\delta}{\delta+\St}}.
\end{equation}
We conducted a simulation of this setup embedded in a protoplanetary disk with a power law background gas density and four dust fluids with Stokes numbers \SIlist{e-3;5e-3;e-2;5e-2}{}. The test domain spans from \SI{10}{\AU} to \SI{30}{\AU} with 200 grid cells, and contains a pressure bump of \SI{2}{\AU} width at \SI{20}{\AU} distance to a solar mass star.
Within the pressure bump, \rev{where the analytical solution applies}, the resulting dust profiles are in excellent agreement with the analytic solutions.
\begin{figure}
    \centering
    \includegraphics{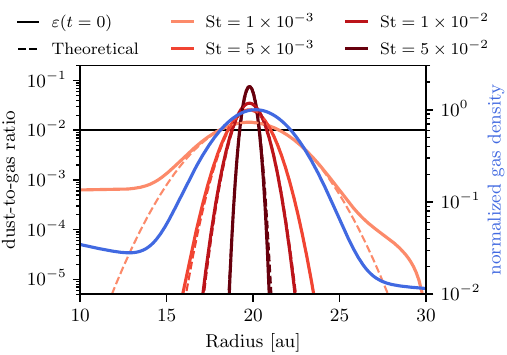}
    \caption{Dust concentration in a steady state radial pressure bump, embedded in a power law density profile. Within the pressure bump, we find excellent agreement between our numerical simulation and the predicted profiles.}
    \label{fig:RadDiff}
\end{figure}

\begin{figure}
    \centering
    \includegraphics{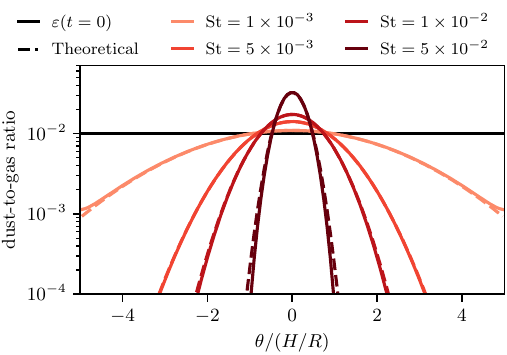}
    \caption{Vertical diffusion test for our passive dust fluid in \pluto{}. We assume a vertically isothermal disk in hydrostatic equilibrium. The resulting dust density distribution agrees well with the analytic prediction.}
    \label{fig:VertDiff}
\end{figure}

As a second test, we set up a two-dimensional, axisymmetric simulation in spherical coordinates. 
We set up a vertically isothermal disk in hydrostatic equilibrium.
Similar to the radial pressure bump, an analytic solution for the settling--mixing equilibrium can be derived which reads
\begin{equation}
    \varepsilon(z)=\varepsilon_0\exp\left(-\frac{z^2}{2H_\mathrm{g}^2}\frac{\St}{\delta}\right)\overset{z\ll R}{\approx} \varepsilon_0\exp\left(-\frac{\vartheta^2}{2(H_\mathrm{g}/R)^2}\frac{\St}{\delta}\right),
\end{equation}
where $\vartheta=\tan(z/R)$.
We find a good agreement between the steady state profiles and the theoretically predicted steady state, as can be seen in \autoref{fig:VertDiff}.

\end{document}